\documentclass[conference]{IEEEtran}

\ifCLASSINFOpdf
\else
\fi

\usepackage[nolist]{acronym}
\usepackage[group-separator={,}, per-mode=symbol]{siunitx}
\usepackage{hyperref}
\usepackage{xspace}
\usepackage{graphicx}
\usepackage{booktabs}
\usepackage{subcaption}
\usepackage{gensymb}
\usepackage{float}
\usepackage{amsmath}
\usepackage{amssymb}
\usepackage{csquotes}

\newcommand{\cf}{\textit{cf}.\xspace}
\newcommand{\etal}[0]{\textit{et~al.}\xspace}
\newcommand{\eg}{\textit{e.g.}\xspace}
\newcommand{\ie}{\textit{i.e.}\xspace}

\newcommand{\schemename}{\textsc{PINsight}\xspace}

\begin{document}
\title{\schemename: Systematic Threat Assessment of Cross-Domain Wi-Fi-based PIN Inference}

\author{\IEEEauthorblockN{Johannes Kortz\IEEEauthorrefmark{1},
Christof Paar\IEEEauthorrefmark{1},
Christian Zenger\IEEEauthorrefmark{2} and Paul Staat\IEEEauthorrefmark{2}
}
\IEEEauthorblockA{\IEEEauthorrefmark{1}Max Planck Institute for Security and Privacy \\ \IEEEauthorrefmark{2}Ruhr University Bochum}
\IEEEauthorblockA{Email: \{johannes.kortz, christof.paar\}@mpi-sp.org,\\\{paul.staat, christian.zenger\}@rub.de}}

\IEEEoverridecommandlockouts

\maketitle

\begin{abstract}
Wi-Fi signals can be repurposed as radar-like sensors, exposing a side channel that adversaries can exploit to infer sensitive information. One particularly concerning example is PIN inference, where the attacker recovers the digits a user types on a keypad by mapping Wi-Fi channel estimations back to individual keystrokes. 
While effective in a fixed setting, where the attacker knows how individual digits map to physical observations, such attacks typically fail once the physical conditions deviate from this setting, \eg, a new room, a different person, or a repositioned device.
The state-of-the-art attack WiKI-Eve tackles this \textit{domain generalization} problem with a deep-learning approach, reporting high PIN inference accuracy regardless of physical conditions -- suggesting a significant real-world threat. However, the threat potential of the attack remains unclear, as isolated success cases cannot substantiate general performance, and no systematic method exists to evaluate performance under unseen conditions.

We close this gap with \schemename, a methodology that separates the effects of changing physical conditions from those of PIN typing itself, enabling a rigorous threat assessment that attributes performance degradation to specific condition changes. \schemename leverages a robotic typing platform that produces highly repeatable keystrokes under systematically varied conditions, such as room and device placement. Using this setup, we record over one million typed digits across more than one thousand controlled combinations of physical conditions, yielding the first benchmark for cross-domain generalization in Wi-Fi PIN inference, which we release publicly. On this benchmark, we revisit the WiKI-Eve attack, address several reproducibility gaps in its evaluation, and construct a stronger variant as an attack baseline. We find that attacks generalize reliably across changes in the background but degrade substantially once the devices are repositioned. We conclude that domain generalization is partially feasible, but that prior results overstate the real-world threat.
\end{abstract}

\IEEEpeerreviewmaketitle

\begin{acronym}[JSONP]
\setlength{\itemsep}{0.2em}
\acro{ATR}{Anti-Tamper Radio}
\acro{BFI}{Beamforming Feedback Information}
\acro{CFR}{Channel Frequency Response}
\acro{CIR}{Channel Impulse Response}
\acro{CSI}{Channel State Information}
\acro{DL}{Deep Learning}
\acro{EM}{electromagnetic}
\acro{ISAC}{Integrated Sensing and Communication}
\acro{NIC}{Network Interface Card}
\acro{OFDM}{Orthogonal Frequency Division Multiplexing}
\acro{PCA}{Principal Component Analysis}
\acro{POS}{Point of Sale}
\acro{RIS}{Reconfigurable Intelligent Surface}
\acro{NDP}{Null Data Packet}
\end{acronym}

\section{Introduction}

Today, billions of \mbox{Wi-Fi} devices are deployed in, \eg, residential, commercial, and public infrastructure~\cite{wifi_device_statistics}. Beyond connectivity, \mbox{Wi-Fi} is increasingly leveraged for high-precision sensing within the emerging \ac{ISAC} paradigm~\cite{liuIntegratedSensingCommunications2022}, enabling standard wireless devices such as smartphones and routers to perform, \eg, human activity recognition, vital sign monitoring, and localization~\cite{maWiFiSensingChannel2019}. However, these capabilities pose a security risk since adversaries can passively eavesdrop on \mbox{Wi-Fi} transmissions and exploit them as a side channel to infer sensitive information about users and their environments. Such adversarial sensing follows the same principles as legitimate sensing: During propagation, radio transmissions interact with the physical environment through reflection, absorption, and scattering caused by surrounding objects and individuals. These physical-layer effects are estimated as an integral part of the \mbox{Wi-Fi} standard~\cite{ieee_80211_phy_standard}, through \ac{CSI} and \ac{BFI}, and therefore can be easily obtained by adversaries~\cite{gringoliFreeYourCSI2019}. This is the foundation of many \mbox{Wi-Fi} sensing attacks~\cite{huPasswordStealingHackingWiFi2023, xiaoLendMeYour2025, zhuTuAlexaWhen2020, banerjeeViolatingPrivacyWalls2014} in which adversaries obtain fine-grained information about victims. 

\textit{PIN inference} is a particularly concerning instance of \mbox{Wi-Fi} sensing attacks that has attracted significant attention from the security community over the last decade~\cite{aliKeystrokeRecognitionUsing2015, chenEchoesFingertipUnveiling2024,chenSilentThiefPassword2024,huPasswordStealingHackingWiFi2023,jinPeriscopeKeystrokeInference2021,shenWiPass1DCNNbasedSmartphone2021,wangMuKIFiMultiPersonKeystroke2024,windtalker,yangWINKWirelessInference2022}. Here, attackers analyze radio signal variations caused by finger movements to recover PINs typed on a victim device, such as a smartphone or a \ac{POS} terminal. Early works~\cite{aliKeystrokeRecognitionUsing2015, windtalker} demonstrated the feasibility of such attacks under fixed physical conditions and in white-box settings, using pattern matching against pre-recorded radio signatures for each digit.

\begin{figure}[t]
    \centering
    \includegraphics[width=1.0\linewidth]{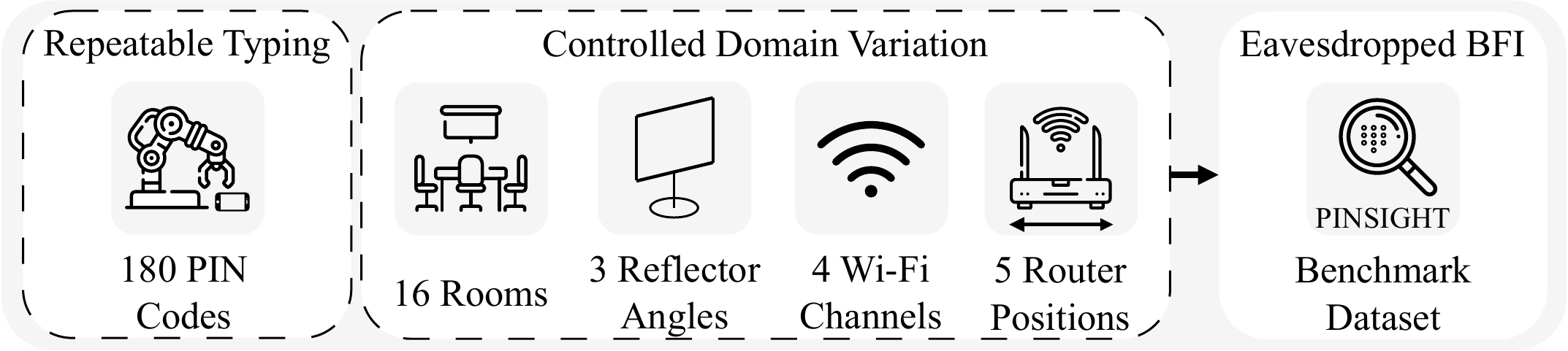}
    \caption{The \schemename dataset for benchmarking cross-domain Wi-Fi-based PIN inference. With repeatable typing and controlled domain variation, it enables targeted ablation studies and rigorous threat assessment.}
    \label{fig:dataset_illustr_fig}
\end{figure}

However, the real-world threat of PIN inference is limited by a lack of generalization across physical conditions. Specifically, the mapping from radio signals to typed digits is highly sensitive to \mbox{\emph{domain shifts}} such as room layout, device placement, and propagation characteristics. A major step toward practical \mbox{Wi-Fi}-based PIN inference under domain shift was presented by Hu~\etal~\cite{huPasswordStealingHackingWiFi2023} at CCS~'23: The WiKI-Eve attack recovers 6-digit PINs within 100 guesses in over \SI{75}{\percent} of cases from eavesdropped \ac{BFI}, and reportedly retains this performance across unseen rooms, users, and victim smartphones. This gain in generalization over prior work~\cite{yangWINKWirelessInference2022} is attributed to a \ac{DL} model trained with an adversarial domain objective. Such a result would be notable, as cross-domain generalization is a well-documented challenge in wireless sensing~\cite{chenCrossDomainWiFiSensing2023}. Were \ac{DL} models able to extract typing-related information from radio propagation across domains, the security and privacy implications would extend well beyond PIN inference to other sensitive signals such as gestures~\cite{abdelnasserWiGestUbiquitousWiFibased2015} and facial expressions~\cite{chenWiFaceFacialExpression2020}.

Despite reported advances in generalization, it remains unclear under which domain shifts PIN inference attacks succeed or fail, and which factors ultimately limit their scalability. WiKI-Eve is evaluated across six rooms, 20 subjects, and six smartphones, without further detail on the experimental conditions, and reported to work throughout, suggesting that domain generalization for PIN inference has been solved. Yet this result reveals little about which shifts the attack actually withstands and may conflate easy and hard ones. Assessing the threat therefore requires a method that measures the effect of individual domain shifts.

To address this gap, we develop \schemename, an experimental approach that, for the first time, enables systematic assessment of cross-domain \mbox{Wi-Fi}-based PIN inference. The idea is to make the sensing event itself invariant. A robotic arm equipped with a phantom hand enters PINs on commodity smartphones with mechanical repeatability, while an eavesdropper collects \ac{BFI} from the \mbox{Wi-Fi} link. Motorized mounts vary the relative position of attacker and victim device, and the entire setup can be moved into different rooms. Because the same digit produces the same physical event under every condition, any change in attack performance can be attributed to the physical condition alone. Using this setup, we record over one million typed digits across more than one thousand controlled combinations of physical conditions, yielding the first benchmark for cross-domain \mbox{Wi-Fi}-based PIN inference. The \schemename dataset and the physical variables are illustrated in~\autoref{fig:dataset_illustr_fig}.

On this benchmark, we evaluate three state-of-the-art attacks, namely WindTalker~\cite{windtalker}, WINK~\cite{yangWINKWirelessInference2022}, and WiKI-Eve~\cite{huPasswordStealingHackingWiFi2023}. Several reproducibility gaps in the original WiKI-Eve evaluation prevent us from reproducing the reported accuracies, so we construct an improved variant, \textit{WiKI-Eve+}, which outperforms all reproductions and baselines we constructed and serves as the basis for our in-depth analysis of domain generalization. Guided by a channel model, we distinguish two classes of domain shift: background changes, which leave the typing signal intact, and encoding changes, which alter it. Across background changes alone, such as a different room or rearranged surrounding objects, generalization works well. However, once the encoding of the typing itself changes, through a different device position or a different \mbox{Wi-Fi} channel, cross-domain performance degrades substantially. WiKI-Eve's \ac{DL} approach thus affords a degree of domain generalization, yet not the robustness that practical domain shifts demand, and the real-world threat is accordingly more constrained than prior results suggest. We publicly release the \schemename benchmark as a controlled basis for reproducible evaluation, inviting the community to test whether stronger cross-domain generalization is achievable.

In summary, we make the following key contributions:

\begin{itemize}
        
    \item We design and implement an experimental method for systematic assessment of cross-domain performance in \mbox{Wi-Fi}-based PIN~inference attacks. 
    
    \item We collect and publish the first large-scale dataset designed specifically for evaluating domain generalization in wireless PIN inference, comprising over one million typed digits across more than one thousand controlled physical conditions.
    
    \item We introduce a two-class taxonomy of physical domain shifts, separating background changes from encoding changes and mapping them to static and mobile attack scenarios. Our analysis shows that state-of-the-art attacks generalize across background changes but degrade under encoding changes.
    
    \item We expose reproducibility failures in WiKI-Eve and show that state-of-the-art domain generalization techniques yield limited cross-domain gains, indicating that prior results overstate the real-world threat.
    
\end{itemize}

\section{Preliminaries}

\subsection{Wi-Fi-based Sensing}

\mbox{Wi-Fi} signals exchanged between a transmitter and a receiver undergo reflection, scattering, absorption, and shadowing. These propagation effects shape the wireless channel, which the receiver estimates from known packet preambles to obtain the \ac{CSI} required for reliable demodulation of the \ac{OFDM} signals. Analogous to radar, wireless sensing leverages \ac{CSI} to infer properties of the physical world from commodity \mbox{Wi-Fi} devices~\cite{zhuCommodityWiFiBasedWireless2024, maWiFiSensingChannel2019}, enabling, for instance, human activity recognition~\cite{zhuTuAlexaWhen2020}, vital signs monitoring~\cite{tewesIRSenabledBreathTracking2022}, and distance estimation~\cite{vasishtDecimeterLevelLocalizationSingle2016}. Access to \ac{CSI} is typically restricted: It resides within the internal signal processing pipeline of \mbox{Wi-Fi} chipsets and requires modified drivers or firmware for extraction~\cite{gringoliFreeYourCSI2019, xiePrecisePowerDelay2015, halperinToolReleaseGathering2011}. %
In contrast, \ac{BFI} enables standard-compliant \mbox{Wi-Fi} sensing on today's unmodified devices. \ac{BFI} is derived from \ac{CSI} at the receiver and fed back to the transmitter for beamforming in unencrypted management frames, which any passive eavesdropper within radio range can observe. \ac{BFI} is a compressed representation of the \ac{CSI} that encodes amplitude and phase across spatial streams and subcarriers. Despite this compression, prior work shows that \ac{BFI}-based sensing achieves performance comparable to \ac{CSI}-based approaches~\cite{haqueBeamSenseRethinkingWireless2023, yi2024bfmsense, haqueWiBFIExtractingIEEE2023, xiaoLendMeYour2025, huPasswordStealingHackingWiFi2023}.

\begin{figure*}[t!]
    \centering
    \includegraphics[width=0.95\linewidth]{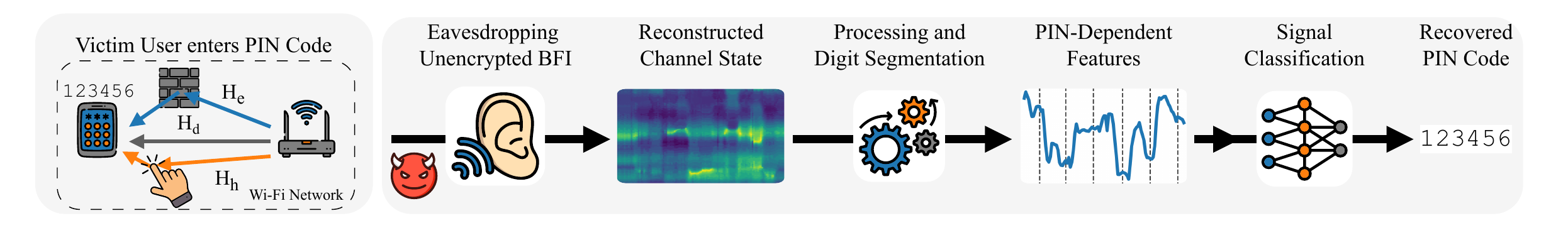}
    \caption{Illustration of signal propagation and Wi-Fi-based PIN inference using BFI in the WiKI-Eve approach~\cite{huPasswordStealingHackingWiFi2023}.}
    \label{fig:attack_concept}
\end{figure*}

\subsection{Threat Model: PIN Inference}
\label{sec:base_threat_model}
We follow the threat model established in prior work~\cite{huPasswordStealingHackingWiFi2023, yangWINKWirelessInference2022, liWhenCSIMeets2016}: A victim uses a smartphone connected to a \mbox{Wi-Fi} network compliant with IEEE~802.11ac or newer and enters a numeric PIN, \eg, to unlock their phone or a sensitive application. The passive attacker Eve is within radio range and eavesdrops on the victim's \ac{BFI} to recover the PIN through this wireless sensing side channel. Because \ac{BFI} is transmitted in plaintext, this requires neither network association nor knowledge of credentials. We assume the attacker knows the standard PIN pad layout of the victim device (\ie, digits arranged as 123 / 456 / 789 / -0-), which is consistent across most smartphone platforms.

As the victim types, finger and hand movements near the smartphone's \mbox{Wi-Fi} antennas modulate the wireless channel, leaving traces in the \ac{BFI} stream. The attacker captures these traces, segments them into per-keystroke intervals, and classifies each segment to recover the PIN (see \autoref{fig:attack_concept}). To train the classifier, the attacker collects \ac{BFI} traces with known PIN entries across domains under their control, but has no knowledge of the victim's domain prior to deployment. The attacker's objective is therefore to learn a mapping from \ac{BFI} to digits that generalizes from these controlled domains to the victim's unseen domain, which defines the cross-domain setting studied in this work.

\subsection{WiKI-Eve Attack}
Whereas prior work relies on per-domain signature matching~\cite{windtalker}, WiKI-Eve is the first to pursue cross-domain digit recognition. It trains a 1D-CNN on eavesdropped BFI with an adversarial domain objective, using a gradient reversal layer~\cite{huPasswordStealingHackingWiFi2023} to suppress domain-specific propagation effects while retaining digit-discriminative information. 
To isolate individual digits, WiKI-Eve segments the \ac{BFI} time series into per-digit windows, each spanning from the preceding keystroke to the following one. For the first and last digits, which lack a neighbor on one side, a fixed offset delimits the open end. Each window therefore captures not only the target keystroke but also the transitions to and from its neighbors. These environment-dependent transition contexts are treated as distinct domains of the same digit, so that factoring them out with the adversarial objective is expected to yield features that transfer to conditions unseen during training.
In a leave-one-out evaluation over six environments, 20 subjects, and six smartphones, WiKI-Eve recovers a 6-digit PIN within 100 guesses in over 75\% of cases when the test room, user, or device is held out during training.

Given these results, the consequences are wide-ranging. \ac{BFI} is available on any \mbox{Wi-Fi} link compliant with IEEE~802.11ac or newer, so a passive eavesdropper could target PIN entry on virtually any modern smartphone, without proximity, device access, or knowledge of the victim's environment. The same capability reaches beyond PINs to the broader class of inference tasks that wireless sensing supports, such as gesture and activity recognition.

\section{Reassessing the Cross-Domain Threat}

The real-world threat of \mbox{Wi-Fi} PIN inference depends on how far the attack generalizes across physical conditions, and WiKI-Eve reports the most extensive cross-domain results to date. Interpreting such results requires knowing which conditions differ between training and deployment, so that performance can be related to the specific factors that changed, such as the room, the device position, or the person typing. We discuss why such an evaluation is difficult to construct and what remains open in the state of the art, motivating the design of \schemename in \autoref{sec:pinsight_main}.

\subsection{The Difficulty of Cross-Domain Assessment}
Assessing cross-domain performance of PIN inference, and of \mbox{Wi-Fi} sensing attacks more broadly, amounts to one question: How does attack performance change when the physical conditions change? Answering this question demands a full characterization of the difference between training and deployment conditions. We identify three obstacles that currently prevent such an assessment.

\paragraph{Physical conditions cannot be controlled as long as the sensing event itself is not repeatable} Human typing varies in finger movement, speed, and hand pose between repetitions, even for the same user. Any change of physical conditions is therefore superimposed with unintended behavioral variation, preventing attribution to either cause.

\paragraph{There is no principled basis for selecting which conditions to vary} Wireless propagation is shaped by the surrounding geometry and material properties, by the position and orientation of the communicating devices, and by the channel in use. These factors span continuous ranges, and their combinations grow rapidly, precluding exhaustive evaluation. Which of them affect PIN inference is not established, leaving no basis for selecting a representative subset.

\paragraph{Cross-domain capabilities are not comparable across attacks} Existing evaluations rely on self-collected datasets recorded under partly documented conditions. Without a shared benchmark, current attacks and future progress cannot be compared with respect to their cross-domain capabilities.
Together, these obstacles lead to a more fundamental problem: For any evaluated domain shift, it remains unknown what actually changed, and consequently how difficult the shift was. When an attack succeeds across a different room, it cannot be determined whether it overcame a substantial change or whether the two rooms were similar. Every reported cross-domain result therefore admits two readings: a generalizing attack or an easy test. Current evaluations cannot distinguish between them.

\subsection{Open Questions in the State of the Art}

The current state-of-the-art attack WiKI-Eve~\cite{huPasswordStealingHackingWiFi2023} reports significant cross-domain capabilities, yet we argue that the implications of the results remain unclear for the following four reasons:

\paragraph{Extent of the domain variation} Results are reported across six environments, six device models, and 20 subjects. Subject-to-AP distances and typing postures are given as ranges, while device and router positions, orientations, and the router model are not specified. The magnitude of the domain shift covered by the evaluation therefore cannot be determined.

\paragraph{Single-axis leave-one-out protocol} Generalization is evaluated by holding out one environment, subject, or device at a time, with the remaining axes present in the training data. The protocol thus measures generalization along a single axis while the others are seen. A joint hold-out -- an unseen subject typing on an unseen device in an unseen environment -- is not evaluated, yet this case corresponds most closely to an attacker targeting a stranger in an unfamiliar location.

\paragraph{Hyperparameter selection without a validation set} The adversarial balance factor, the segmentation parameters, and the sparse-recovery sampling rate are reported as empirically determined on a single 70/30 train-test split, with no held-out validation set. Selecting these values on the test data would bias the reported accuracies, and the protocol does not rule this out.

\paragraph{Mechanisms of generalization} Robustness across environments is attributed to the diffraction pattern of the hand at the phone body, described as largely invariant to the surroundings. We will show that small changes in the relative device position already cause substantial \ac{BFI} variation (\autoref{sec:setup_validation}). The benefit of the adversarial objective is supported by a t-SNE projection of the learned features rather than a cross-domain measurement, with no ablation against an otherwise identical non-adversarial model. Our own ablation finds no consistent advantage over standard training (\autoref{sec:adaptation_methods}).

Resolving these questions requires an evaluation in which physical conditions are varied in a controlled manner and the sensing event is repeatable. \schemename provides such an evaluation.

\section{\schemename: Theoretical Model and Experimental Setup}
\label{sec:pinsight_main}

We introduce \schemename, a methodology for systematically assessing the threat of \mbox{Wi-Fi}-based PIN inference. At its core is a theoretical model that structures the unbounded space of physical conditions and informs the experimental design. Combined with robotic PIN entry, which makes the sensing event reproducible, this allows any change in attack performance to be attributed to the domain rather than to the way the victim typed. The resulting benchmark and its evaluation methodology follow in \autoref{sec:benchmark_dataset}.

\subsection{Theoretical Model}
\label{sec:theoretical_model}

Domain shifts in PIN inference break down into two fundamental classes of physical conditions, which in real-world scenarios mostly occur in combination: changes that leave the typing signal intact and alter only the background, and changes that cause the typing signal itself to vary. This model categorizes all possible physical domain shifts into one of two classes to set the axes of our experiment.

\paragraph{Channel decomposition} The attack rests on the victim's hand affecting the propagation of \mbox{Wi-Fi} signals. Following \autoref{fig:attack_concept}, the reciprocal complex-valued channel between smartphone and router on a given \ac{OFDM} subcarrier and antenna pair at time $t$ splits into three components:
\begin{equation}
    H(t) = H_{\mathrm{d}}(t) + H_{\mathrm{e}}(t) + H_{\mathrm{h}}(t), \label{eq:channel_decomp}
\end{equation}
the direct paths that interact with neither the environment nor the hand, the paths that interact with the environment but not the hand, and the paths that interact with the hand. The hand-dependent component $H_{\mathrm{h}}(t) = \sum_{\ell \in \mathcal{H}(t)} h_{\ell}^{(\mathrm{h})}(t)$ is the sum over the set $\mathcal{H}(t)$ of all hand-interacting paths, each with time-varying gain $h_{\ell}^{(\mathrm{h})}(t)$ capturing the propagation to and from the hand along path $\ell$. Only $H_{\mathrm{h}}(t)$ carries the typing signal, observed by the attacker as \ac{CSI} or \ac{BFI}.

Each hand path traverses three stages, from the smartphone to the hand, through the hand, and from the hand to the router. This matches a bistatic scattering model with the hand as a passive scatterer between two separated endpoints~\cite{cherniakovBistaticRadarEmerging2008},
\begin{equation}
    h_{\ell}^{(\mathrm{h})}(t) = g_{\ell}^{\mathrm{s} \to \mathrm{h}}(t) \cdot s_{\ell}(t) \cdot g_{\ell}^{\mathrm{h} \to \mathrm{r}}(t), \label{eq:hand_decomp}
\end{equation}
with the segment gains $g_{\ell}^{\mathrm{s} \to \mathrm{h}}(t)$ and $g_{\ell}^{\mathrm{h} \to \mathrm{r}}(t)$ and the scattering response $s_{\ell}(t)$, set by hand pose, finger position, and tissue properties. 

\paragraph{Background changes} Moving furniture, relocating objects, or switching rooms changes
$H_{\mathrm{e}}(t)$ and leaves the remaining terms of
\autoref{eq:channel_decomp} intact. $H_{\mathrm{h}}(t)$ is untouched, so every digit
produces the same response against a different background, and a model that
suppresses the background still recognizes the digit. This is a classical case of additive interference: the attacker's task is to suppress a changed background. It captures scenarios in which devices are stationary, such as \ac{POS} terminals, and only the surroundings differ between training and attack.

\paragraph{Encoding changes}
Changing the relative position of smartphone and router, the \mbox{Wi-Fi}
channel, or the typing hand reaches into \autoref{eq:hand_decomp} itself. Such changes either reconfigure the segment gains $g_{\ell}^{\mathrm{s} \to \mathrm{h}}(t)$ and $g_{\ell}^{\mathrm{h} \to \mathrm{r}}(t)$ or the scattering response $s_{\ell}(t)$. Together,
these factors define how typing is written into the channel, which we call the \emph{encoding}. Every encoding change is accompanied by a background change, while a background change leaves the encoding untouched. This class captures mobile scenarios, such as handheld smartphones, where device positions vary between training and attack.

\paragraph{Towards structured domain shifts}
This taxonomy enables a structured exploration of domain shifts of varying
extent. We first vary the background in graded steps, and subsequently the encoding. For background changes, relative device positions, the \mbox{Wi-Fi} channel, and the typing must remain strictly fixed while the surroundings vary. Any change beyond that, \eg, moving the router or switching the \mbox{Wi-Fi} channel, yields an encoding change.

\subsection{Experimental Setup}
In the following, we outline the experimental setup for automated PIN typing and controlled domain shift. We begin with the elements \schemename shares with prior PIN inference attacks and then detail the components that we add.

\paragraph{Eavesdropping setup}
Unless indicated otherwise, we use an Apple iPhone~13 as the victim PIN entry device, one of the smartphones on which WiKI-Eve was evaluated~\cite{huPasswordStealingHackingWiFi2023}. It is connected to an IEEE~802.11ac \mbox{Wi-Fi} network on a Netgear Nighthawk~RAX50 router at a distance of \SI{1.5}{\m}, with a channel bandwidth of \SI{80}{\MHz}. The router periodically transmits \ac{NDP} sounding frames at a rate of approximately~\SI{18}{\Hz}, triggering the phone to report \ac{BFI} feedback. With four router antennas and two phone antennas, each \ac{BFI} report comprises a $2 \times 4$ spatial channel matrix for each of the $234$~subcarriers. The attacker employs a Lenovo ThinkPad~T14~Gen~2i equipped with an Intel~AX201 \mbox{Wi-Fi} \ac{NIC} configured in monitor mode to eavesdrop on the \ac{BFI} reports transmitted by the smartphone. To simplify the experimental setup, the laptop is placed on the typing platform, yet the attacker can capture \ac{BFI} from any location within \mbox{Wi-Fi} range. We use TShark to capture and filter \ac{BFI} reports and build on the approach in~\cite{haqueWiBFIExtractingIEEE2023} to reconstruct complex-valued beamforming feedback matrices from \ac{BFI}. 

\paragraph{Robotic typing platform}

\schemename decouples PIN entry from domain shifts at both the background and encoding level by automating it with a robotic arm, producing identical
keystroke sequences across physical conditions. Holding the sensing event
invariant is what lets us attribute any change in attack performance to the
domain shift under study rather than to typing behavior. To this end, we mount
the smartphone, robotic typing mechanism, and \mbox{Wi-Fi} router on a shared
wooden base that preserves their relative geometry while the station itself is
relocated or reconfigured.

\begin{figure}[t]
    \centering
    \sbox0{\includegraphics[width=0.7\linewidth]{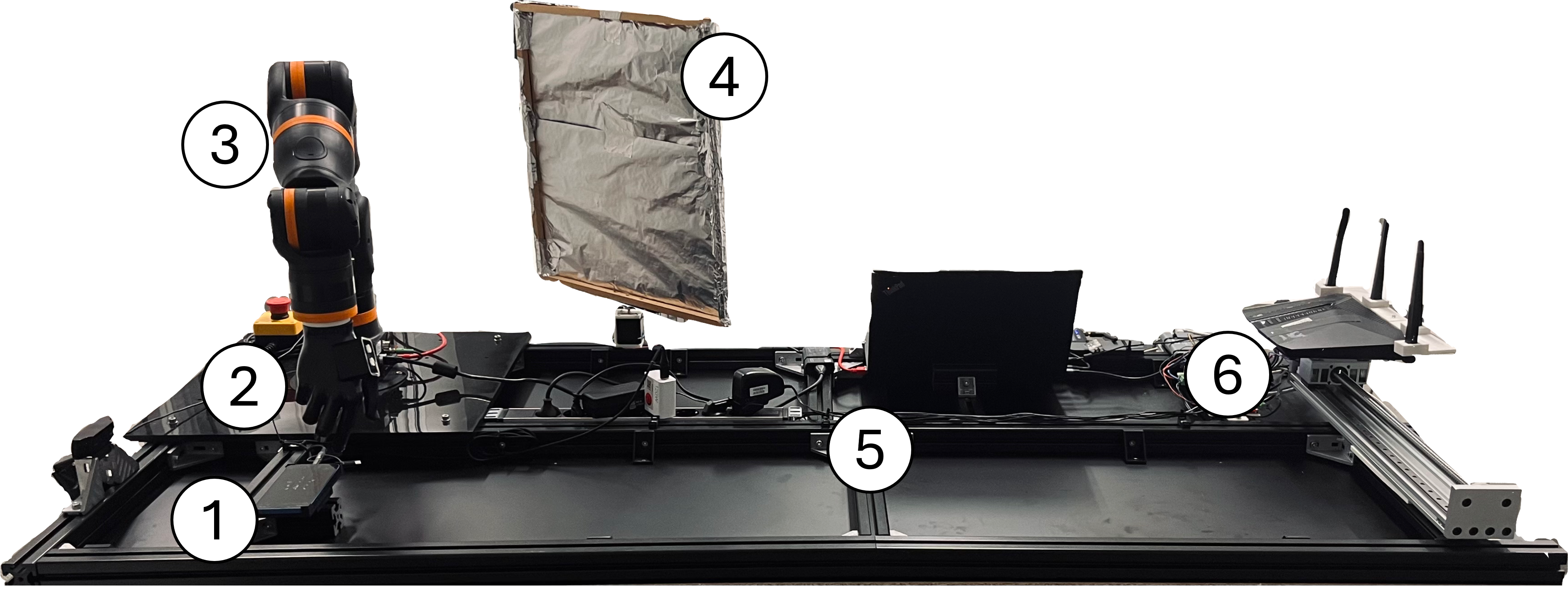}}%
    \subcaptionbox{Overview of the platform.\label{fig:setup_overview}}{%
        \usebox0%
    }\hfill
    \subcaptionbox{Detail view.\label{fig:setup_detail}}{%
        \includegraphics[height=\ht0]{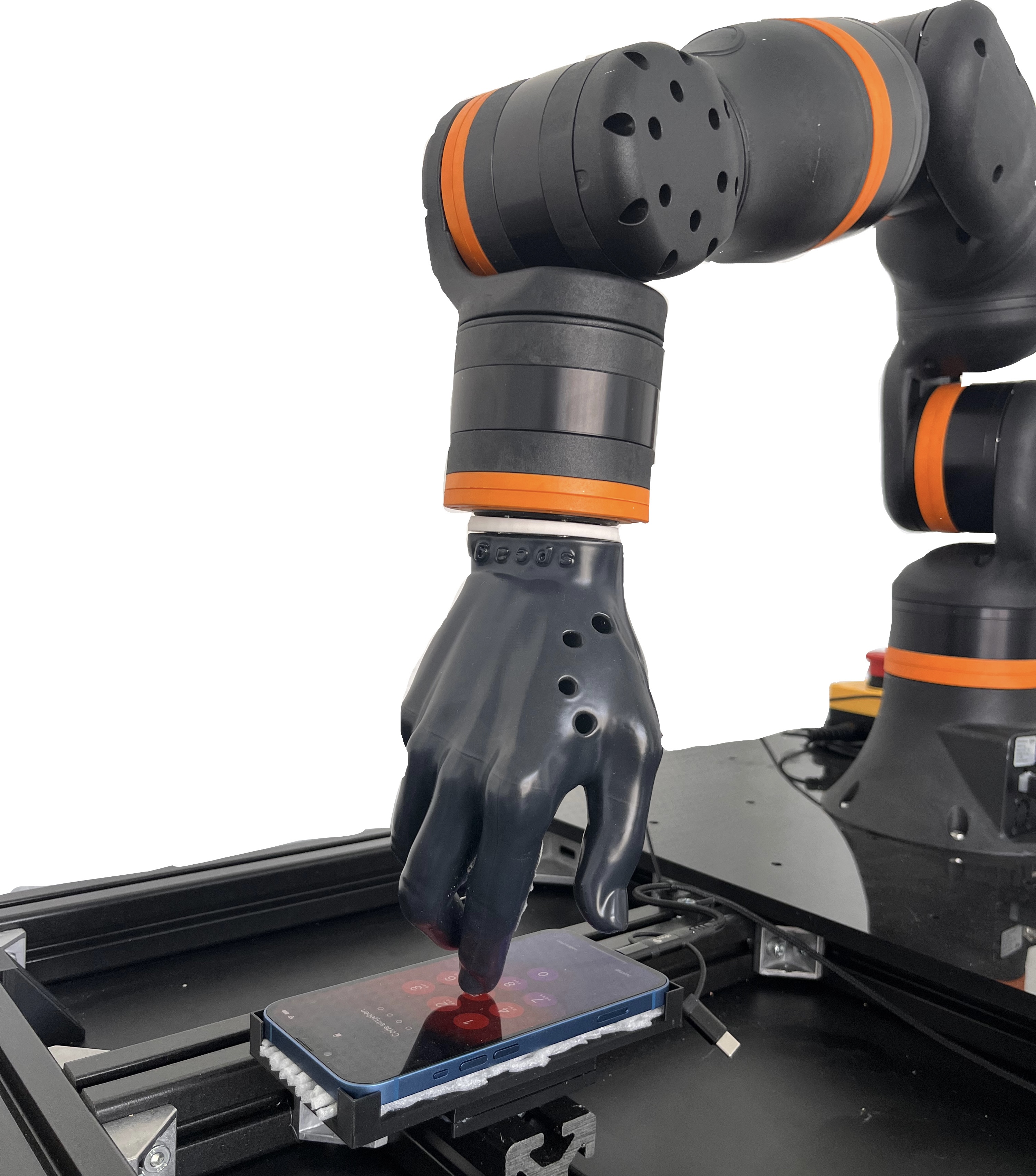}%
    }
    \caption{Experimental platform for repeatable PIN entry under controlled domain shift, comprising the victim smartphone (1), an EM phantom hand (2) on a robotic arm (3), a rotatable reflector (4), the attacker laptop (5), and a Wi-Fi router on a linear axis (6). The platform can be relocated across rooms while maintaining fixed relative geometry.}
    \label{fig:experimental_setup}
\end{figure}

To type repeatable PIN sequences, we employ an igus REBEL-6DOF-03 six-axis robotic arm with \SI{1}{\mm} positioning precision. Its predominantly
non-metallic construction minimizes its own imprint on the wireless channel, avoiding unrealistically strong distortion. To the arm's tip we attach an \ac{EM} phantom hand from SPEAG, engineered to replicate the \ac{EM} properties of a human hand for over-the-air smartphone antenna
testing~\cite{hand_phantom_effects_2009, christReflectionPropertiesHuman2021} (\autoref{fig:experimental_setup}).

With the index finger pointing downward, the phantom hand executes keystrokes on
the iPhone~13 standard PIN pad. Each entry begins and ends at a randomized
position \SI{10}{\mm} above the phone, preventing implicit leakage from fixed
entry or exit trajectories. Between consecutive digits, the hand rises to at most
\SI{20}{\mm} before descending to lightly contact the center of the target
digit. At maximum speed this yields roughly \SI{0.8}{\s} per digit, about
$14.4$~\ac{BFI} samples at \SI{18}{\Hz}. We resample the \ac{BFI} time series to
compensate for non-linearities in the arm's distance-time relationship,
enforcing constant hand speed.

Background-type domain shift is induced by ($i$)~relocating the station across \num{16}~rooms (see~\autoref{tab:rooms} in the Appendix) and ($ii$)~rotating an aluminum-foil reflector of size \SI{0.5}{\m}~$\times$~\SI{0.5}{\m} into three discrete positions (in \SI{45}{\degree} steps) using a stepper motor. Encoding-type domain shift is introduced through ($i$)~router displacement along five positions in \SI{15}{\cm} steps on a \SI{60}{\cm} motorized linear axis and ($ii$)~Wi-Fi channel changes (across the channels \num{44}, \num{56}, \num{104}, and \num{120}).

\subsection{Validation Experiments}
\label{sec:setup_validation}

\begin{figure}[t]
\centering
\begin{subfigure}{0.225\textwidth}
    \includegraphics[width=1.0\linewidth]{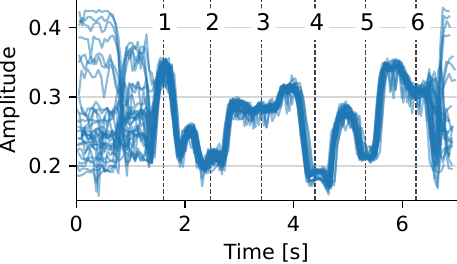}
    \caption{Robotic PIN \texttt{123456}}
    \label{fig:spectrum}
\end{subfigure}
\quad
\begin{subfigure}{0.225\textwidth}
    \includegraphics[width=1.0\linewidth]{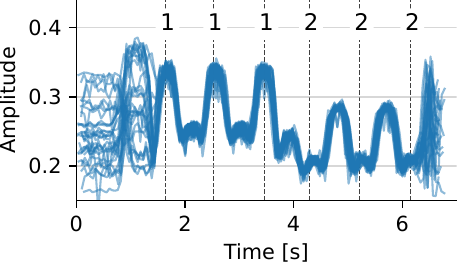}
    \caption{Robotic PIN \texttt{111222}}
    \label{fig:time}
\end{subfigure}
\\
\begin{subfigure}{0.225\textwidth}
    \includegraphics[width=1.0\linewidth]{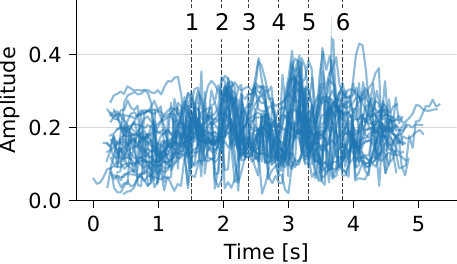}
    \caption{Human PIN \texttt{123456}}
    \label{fig:humanspectrum}
\end{subfigure}
\quad
\begin{subfigure}{0.225\textwidth}
    \includegraphics[width=1.0\linewidth]{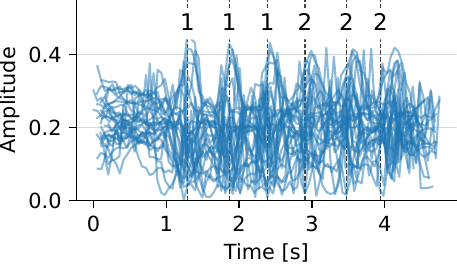}
    \caption{Human PIN \texttt{111222}}
    \label{fig:humantime}
\end{subfigure}
\caption{Amplitude time series of BFI averaged over all subcarrier frequencies for a single spatial stream. The same PIN repeated 30~times.}
\label{fig:repetition}
\end{figure}

\paragraph{Repeatable robotic typing}
To verify the repeatability of the setup, we conduct an experiment in
which the robotic arm repeatedly enters the PIN sequences \texttt{123456} and \texttt{111222} while recording \ac{BFI}. \autoref{fig:repetition} shows the
subcarrier-averaged \ac{BFI} time series of a single spatial stream, with
keystrokes marked by vertical dashed lines. In \autoref{fig:spectrum} and
\autoref{fig:time} the traces align closely across repetitions of the same
sequence, confirming that the robotic setup produces repeatable channel
observations. The overlapping \texttt{1-2} segment shared by both PINs yields consistent \ac{BFI} patterns across sequences, while the variation before and after typing stems from the randomized start and end positions. We further confirm that the observed effect on the \ac{BFI} traces originates from the hand and not from the robotic arm itself (see~\autoref{fig:phantom_hand_effect} in the Appendix). For comparison, we repeat the experiment with a human typist instead of the robotic arm to enter the same sequences in the same physical domain. Although the typist was instructed to type consistently, \autoref{fig:humanspectrum} and \autoref{fig:humantime} show high variation
across repeated entries: while repeated digits exhibit some similarity, the behavioral noise of human typing prohibits controlled cross-domain evaluation.

\autoref{fig:repetition} illustrates a key property of our \schemename setup: Robotic PIN entry removes the behavioral variability of human typing,
minimizing variation in finger movement, speed, and hand pose. This lets us
examine the effect of domain shifts in isolation, though the idealized typing does not reflect true human behavior. Note that this grants the attacker an advantage, since idealized typing is arguably simpler to infer.

\begin{figure}[t]
\centering
\begin{subfigure}[b]{0.47\textwidth}
    \includegraphics[width=1.0\linewidth]{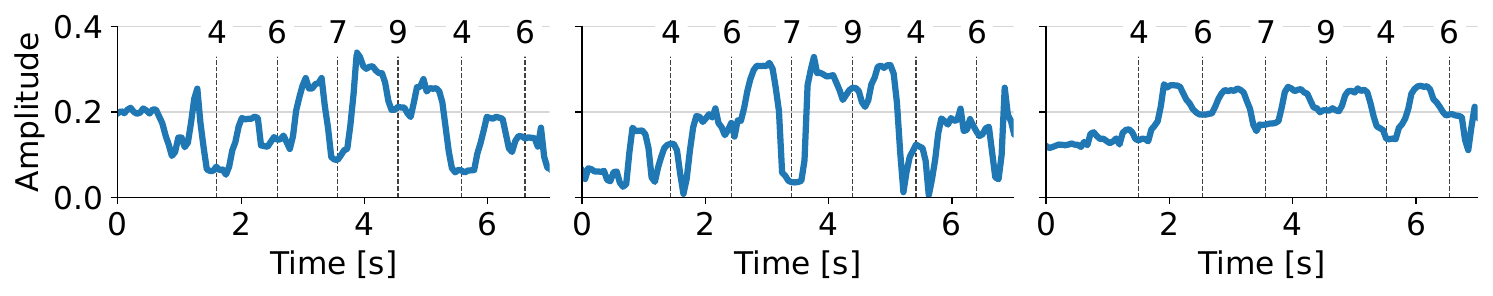}
    \caption{Left to right: rooms 3, 4, and 10 with router position 1.}
    \label{fig:10a}
\end{subfigure}
\\
\begin{subfigure}[b]{0.47\textwidth}
    \includegraphics[width=1.0\linewidth]{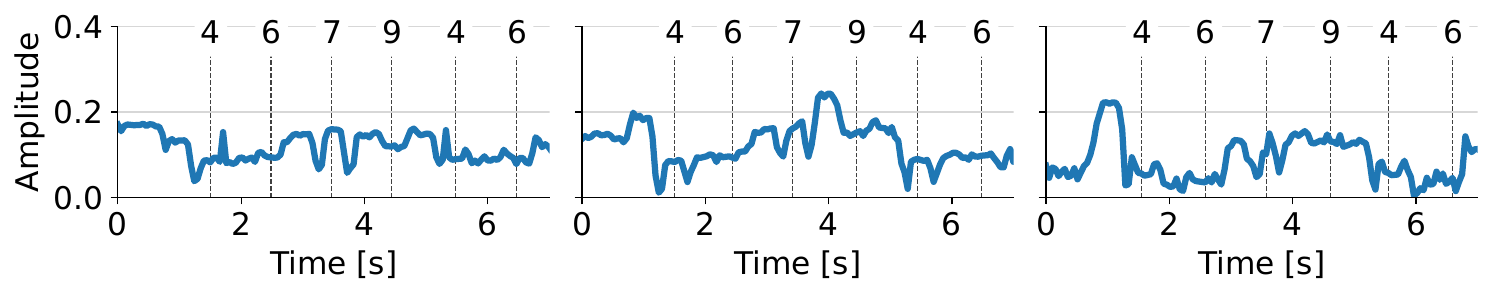}
    \caption{Left to right: rooms 3, 4, and 10 with router position 3.}
    \label{fig:10b}
\end{subfigure}
\caption{Amplitude time series of BFI averaged over all subcarrier frequencies for a single spatial stream. The PIN \texttt{467946} is typed across three rooms and two router positions. Notably, rooms~3 and~4 exhibit similar patterns.}
\label{fig:domain_examples}
\end{figure}
\paragraph{Controlled domain variation}
An example of the effect of physical domain shift on the \ac{BFI} data observed by the attacker is shown in \autoref{fig:domain_examples}. Here, we again plot the subcarrier average over one spatial channel of \ac{BFI} for the same PIN being typed in three different rooms and for two different router positions. The controlled changes in room and router placement visibly alter the observed signal, confirming that our setup produces measurably distinct domains for identical typing input. Crucially, these domain changes do not all have the same impact. Under router position~1, rooms~3 and~4 exhibit similar amplitude dynamics, whereas room~10 differs markedly. Moving the router to position~3 alters the traces in all three rooms, changing the keystroke signatures in each. 

\begin{figure}[t]
\centering
\sbox0{\includegraphics[width=0.12\columnwidth]{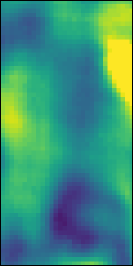}}%
\captionsetup[subfigure]{font=scriptsize}
\subcaptionbox{\SI{0}{\cm}\\(baseline)\label{fig:grida}}[0.14\columnwidth]{\usebox0}\hfill
\subcaptionbox{\SI{1}{\cm}\\$\rho{=}0.95$\label{fig:gridb}}[0.14\columnwidth]{%
    \includegraphics[width=0.12\columnwidth]{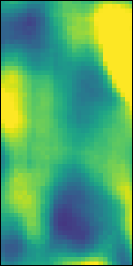}}\hfill
\subcaptionbox{\SI{2.5}{\cm}\\$\rho{=}0.83$\label{fig:gridc}}[0.14\columnwidth]{%
    \includegraphics[width=0.12\columnwidth]{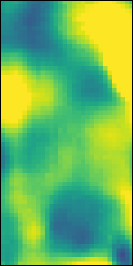}}\hfill
\subcaptionbox{\SI{5}{\cm}\\$\rho{=}0.56$\label{fig:gridd}}[0.14\columnwidth]{%
    \includegraphics[width=0.12\columnwidth]{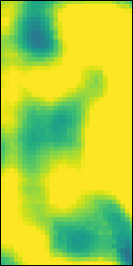}}\hfill
\subcaptionbox{\SI{7.5}{\cm}\\$\rho{=}0.51$\label{fig:gride}}[0.14\columnwidth]{%
    \includegraphics[width=0.12\columnwidth]{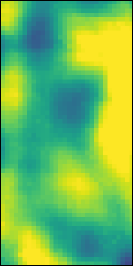}}\hfill
\subcaptionbox{\SI{15}{\cm}\\$\rho{=}0.13$\label{fig:gridf}}[0.14\columnwidth]{%
    \includegraphics[width=0.12\columnwidth]{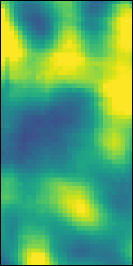}}\hfill
\raisebox{0pt}[\ht0][0pt]{%
    \includegraphics[height=\ht0]{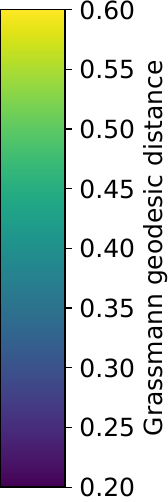}}
\caption{Sensitivity of BFI to vertical hand displacement, measured across a grid of $30 \times 60$ positions over the smartphone, for six router-position offsets from
baseline. We report the Pearson correlation $\rho$ of the
sensitivity map to the baseline.}
\label{fig:grids}
\end{figure}

To investigate the effect of physical domain shift on the hand-to-\ac{BFI} mapping more comprehensively, we alter router placement, \ie, encoding-type variation introduced in \autoref{sec:theoretical_model}. In particular, we characterize sensitivity to vertical hand motion. Across a grid of $30 \times 60$ positions over the smartphone, we record \ac{BFI} with the hand on the phone and again with the hand raised \SI{20}{\mm}, and measure per-position sensitivity using a distance metric between the two \ac{BFI} measurements. As the same physical channel can map to different \ac{BFI} matrices, we rely on the Grassmann geodesic distance, which compares the subspaces they span. The resulting heatmaps (\autoref{fig:grids}) capture the spatial effect of vertical displacement on \ac{BFI}. Repeating this procedure while displacing the \mbox{Wi-Fi} router along the linear axis reveals that the hand-effect patterns become increasingly dissimilar from the baseline. The Pearson correlation to the baseline heatmap remains high at a displacement of \SI{1}{\cm} ($\rho = 0.95$) but falls to $\rho = 0.56$ by \SI{5}{\cm} and to $\rho = 0.13$ at \SI{15}{\cm}, where the patterns are effectively unrelated. This highlights the significant effect of domain shifts on \ac{BFI}: Even small changes in device position substantially reshape the hand-to-\ac{BFI} mapping.

Overall, this result stands in contrast to the robustness argument put forward for WiKI-Eve. Beyond adversarial learning, the attack's cross-domain performance across environments is attributed to ``the diffraction pattern around the phone body that are rarely influenced by environment-specific interference''~\cite{huPasswordStealingHackingWiFi2023}, further motivating a systematic reassessment.

\section{Large-Scale Domain Shift Dataset}
\label{sec:benchmark_dataset}

We use the setup described above to collect the \schemename dataset for
evaluating domain shift in \mbox{Wi-Fi}-based PIN inference attacks. The main
benchmark spans \num{960} controlled domains with \num{180} PINs each. With the
targeted transfer conditions of \autoref{sec:transfer_dataset}, \schemename
comprises over one thousand domains and more than one million typed digits.

\subsection{Benchmark Dataset}
The dataset spans 960 domains across 16 rooms, each covering all combinations of five router positions, three reflector angles, and four \mbox{Wi-Fi} channels, \ie, the background-type and encoding-type shift axes introduced in \autoref{sec:theoretical_model}. The key variables are illustrated in \autoref{fig:dataset_illustr_fig}. To our knowledge, it is the first large-scale dataset explicitly designed to isolate these shift axes in PIN inference.

For each domain, we collect 180 six-digit PINs in two groups. The first group is 100 PINs drawn randomly per domain. The second is 80 fixed PINs reused verbatim across all domains: 60 drawn randomly once and reused, and 20 selected manually to include the most common PINs alongside deliberate repetitions for assessing typing repeatability. The fixed group enables direct cross-domain comparison of identical input sequences.

The \schemename dataset is publicly available at~\cite{pinsight_repo} and comprises \num{1036800} typed digits across \num{172800} six-digit PINs. For each PIN, we provide the raw eavesdropped \ac{BFI} frames, parsed complex-valued channel matrices, and feature time series. Every \ac{BFI} sequence is accompanied by the corresponding robotic hand positions and labeled with the typed digit sequence and domain metadata, enabling reproducible evaluation of domain generalization as studied in this paper.

\subsection{Transfer Dataset}
\label{sec:transfer_dataset}
To evaluate cross-domain transfer beyond the parameter set covered by the main typing platform, we collect a complementary transfer dataset. For each transfer condition, we record \num{80} six-digit PINs. The conditions cover a range of encoding-type and background-type shifts, introducing additional changes to the environment, the router positioning, the phone, and the typing. This dataset enables targeted assessment of attack generalization along individual physical axes, complementing the systematic coverage of the main \schemename dataset.

\subsection{Benchmarking Domain Generalization}
\label{sec:benchmark}
To use \schemename as a domain-generalization benchmark, we need a protocol that separates the four domain factors (room (R), reflector angle (A), \mbox{Wi-Fi} channel (W), and router position (P)) into training, validation, and test sets. We adopt a leave-out strategy that excludes specific domain factors from training and validation. From the six possible factor pairs we select four (\autoref{tab:second-order-leave-out}).
The unseen room-and-position pair is of primary interest, as it reflects an attacker who can anticipate neither the victim's environment nor the device geometry. The remaining combinations provide complementary views on the relative difficulty of different domain shifts.

Each selected pair defines a \emph{second-order leave-out}: one instance of two factors is held out for testing, while all instances of the other factors appear in training, validation, and test. A single such split supports three tests. Two \emph{first-order} tests evaluate each held-out factor on its own, pairing its unseen instance with all seen combinations of the other factors. Aggregating over those combinations gives an estimate of transfer along that one axis. The \emph{second-order} test combines both held-out instances, measuring a simultaneous shift along two factors at once.

\begin{table}[t]
\centering
\caption{Second-order leave-out evaluation splits. For each domain-shift type,
we report the number of seen (S) and unseen (U) instances.}
\label{tab:second-order-leave-out}
\setlength{\tabcolsep}{4.5pt}
\newcommand{\su}[2]{\makebox[2.2em][c]{#1\hspace{0.5em}#2}}
\begin{tabular}{llccccc}
\toprule
ID & Leave-Out & Room & Pos. & Wi-Fi & Refl. & Seen \\
   &           & \su{S}{U} & \su{S}{U} & \su{S}{U} & \su{S}{U} & Domains \\
\midrule
RP & Room, Pos.      & \su{15}{\textbf{1}} & \su{4}{\textbf{1}}  & \su{4}{--}           & \su{3}{--}           & 720 \\
RW & Room, Wi-Fi         & \su{15}{\textbf{1}} & \su{5}{--}           & \su{3}{\textbf{1}}  & \su{3}{--}           & 675 \\
RA & Room, Refl.     & \su{15}{\textbf{1}} & \su{5}{--}           & \su{4}{--}           & \su{2}{\textbf{1}}  & 600 \\
AP & Refl., Pos. & \su{16}{--}         & \su{4}{\textbf{1}}  & \su{4}{--}           & \su{2}{\textbf{1}}  & 512 \\
\bottomrule
\end{tabular}
\end{table}

For model selection (hyperparameter tuning and early stopping), we follow~\cite{gulrajani2020searchlostdomaingeneralization} and use a \emph{training-domain validation set}. Specifically, we pool data from all seen domains and split it at the class level, using \SI{80}{\percent} of digits for training and \SI{20}{\percent} for validation. This approach has been shown to outperform leave-one-domain-out cross-validation while maximizing the data available for training~\cite{gulrajani2020searchlostdomaingeneralization}. 

To obtain a representative estimate of generalization along a given leave-out test, cross-validation is necessary: we iterate over multiple instances of the held-out factor(s) and report the mean and standard deviation across these folds. This per-instance repetition is essential, as we show in \autoref{sec:deepdive} that even for the same type of transfer, results can vary substantially depending on which instance is held out.

\subsection{Evaluation Protocol}

Beyond the standard threat model presented in \autoref{sec:base_threat_model}, we make additional simplifying assumptions to isolate the \ac{BFI}-to-digit inference problem: we assume the attacker can identify the time window of each keystroke, and is provided with fully sampled, non-sparse \ac{BFI}, known device identity, and exact keystroke timing for perfect digit segmentation. This removes confounds such as traffic capture, device identification, and segmentation errors, so that measured domain effects reflect the inference task rather than pipeline artifacts.

To assess attack success, we use \emph{Top-100 accuracy}, a metric commonly adopted in prior work~\cite{yangWINKWirelessInference2022, huPasswordStealingHackingWiFi2023, windtalker}. It measures the fraction of test PINs for which the correct six-digit PIN appears among the 100 most likely candidates, ranked by the product of per-digit probabilities. Inclusion in the Top-100 reduces the search space from $10^6$ to 100 candidates, corresponding to a reduction factor of $10^4$, assuming the attacker can attempt multiple guesses.

\section{Evaluating State-of-the-Art PIN Inference}
\label{chap:attack_methods}

\schemename enables a fair comparison of PIN inference methods in the cross-domain setting. Using this benchmark and the leave-out protocol of \autoref{sec:benchmark}, we evaluate, for the first time, different attacks under identical, controlled domain shifts. We proceed in four steps. $(i)$ We examine the WiKI-Eve attack and find that it cannot be reproduced faithfully from the available material. $(ii)$ We therefore compare several feature representations and architectures to establish the strongest realization of the WiKI-Eve approach we can construct. $(iii)$ We assess the role of adversarial learning, which WiKI-Eve presents as the mechanism behind its cross-domain performance. $(iv)$ Finally, we benchmark our strongest WiKI-Eve realization against the other state-of-the-art methods, WINK and WindTalker.

\subsection{Barriers to Reproducing WiKI-Eve}
Reproducing the original WiKI-Eve~\cite{huPasswordStealingHackingWiFi2023}
attack required filling in numerous unspecified implementation details. The artifact repository accompanying WiKI-Eve~\cite{github_wiki_eve} provides \ac{BFI} parsing and a TCN-based sparse-recovery module for upsampling sparse \ac{BFI} recordings. Its training scripts operate on respiration, gesture, and activity data and appear to import dataloaders from a different work. Crucially, it contains no adversarial learning pipeline, model architecture, dataset, or
trained weights, and thus does not support reproduction of the reported PIN inference results. Our requests for clarification remained unanswered. From the paper alone, we were unable to reconstruct the attack implementation either, for the following two reasons:

\paragraph{Ambiguous model architecture} Figure~10 of~\cite{huPasswordStealingHackingWiFi2023}
depicts only a coarse architecture: Hyperparameters such as the number of filters per convolutional layer, the kernel sizes, and the exact number of layers remain unspecified.

\paragraph{Underspecified signal representation} \ac{BFI} is inherently a tensor over time, spanning subcarrier and spatial-stream dimensions, yet WiKI-Eve classifies digits from a one-dimensional time series without specifying how the tensor is collapsed to that series. Figure~3 of~\cite{huPasswordStealingHackingWiFi2023} shows \ac{BFI} amplitude traces of keystrokes with values up to~50, whereas reconstructed \ac{BFI} has a maximum amplitude of one by construction, suggesting that WiKI-Eve applies undocumented processing to the \ac{BFI} before classification.

\subsection{WiKI-Eve+}

\begin{figure}[t]
    \centering
    \includegraphics[width=0.65\linewidth]{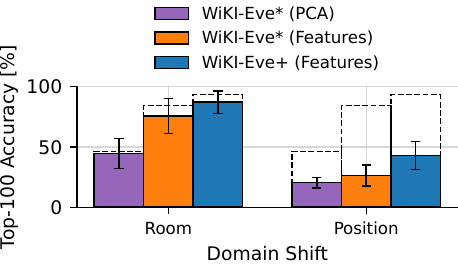}
    \caption{Comparison of input representations and architectures. Hand-crafted features outperform \ac{PCA}, and WiKI-Eve+ outperforms WiKI-Eve*. Error bars show the standard deviation across leave-one-out folds and dashed lines show the average in-domain validation accuracy.}
    \label{fig:wiki_comparison}
\end{figure}

We address both reproducibility gaps by constructing and comparing several variants, isolating the effect of the signal representation and of the model architecture. 

We begin with the signal representation. As direct classification on raw \ac{BFI} was ineffective in our experiments, we implement two pipelines. The first applies \ac{PCA} to extract a single value per spatial stream, collapsing the subcarrier dimension. This is a common step in the wireless-sensing literature~\cite{palipanaFallDeFiUbiquitousFall2018}, as neighboring subcarriers carry redundant information that \ac{PCA} removes while retaining the directions of maximum variance. The second extracts 17 hand-crafted feature groups (detailed in~\autoref{tab:features} in the Appendix) that capture temporal
keystroke signatures while preserving the spatial and spectral structure of \ac{BFI}, reducing each time sample from 1872 to 134 dimensions.

We then address the ambiguous model architecture with two variants.
\mbox{\textbf{WiKI-Eve*}} is a best-effort reproduction of the architecture in
Figure~10 of~\cite{huPasswordStealingHackingWiFi2023}, whose feature extractor
combines a 1D convolutional layer with adaptive average pooling. This
reproduction reaches only weak PIN inference accuracy even in-domain, which
motivates a higher-capacity design. \mbox{\textbf{WiKI-Eve+}} therefore replaces the
extractor with time-local feature mixing and a latent-embedding layer
(\cf~\autoref{fig:model_arch} in the Appendix)\footnote{We also evaluated a WiKI-Eve+ variant with a Transformer encoder in
place of the convolutional feature extractor (\autoref{fig:trafo} in the
Appendix). It underperforms the convolutional model, indicating that our results
are not an artifact of the chosen architecture.}, giving the model more capacity to combine the
hand-crafted features into a domain-invariant representation. The resulting
model has roughly 3M parameters and is regularized with dropout and weight
decay. The two variants also differ in their input construction: where WiKI-Eve*
pairs two concatenated samples (a digit with its surrounding transitions) and predicts whether they share a domain, WiKI-Eve+ operates on single samples, predicting whether a sample originates from a fixed reference domain or differs from it. This avoids the asymmetry between paired training and duplicated-sample inference and follows the canonical DANN formulation~\cite{ganin15}, from which the original WiKI-Eve work departs
without a stated rationale.

\autoref{fig:wiki_comparison} compares these variants. Under both room and
position shift, the hand-crafted-feature variant of WiKI-Eve* outperforms the
\ac{PCA} variant, so we adopt the hand-crafted features for all subsequent
evaluations. WiKI-Eve+ raises performance further, surpassing both WiKI-Eve*
variants. As it outperforms every faithful reproduction we could construct, we
treat WiKI-Eve+ as the strongest realization of the attack and use it throughout
the remainder of the paper to probe the cross-domain threat of PIN inference.

For the final WiKI-Eve+ variant we make one further change. Because the
adversarial objective can only enforce invariance to what the discriminator
distinguishes, we derive domain labels from the physical factors along which we
evaluate generalization (room, reflector angle, \mbox{Wi-Fi} channel, and
router position) rather than from neighboring-digit context
as in~\cite{huPasswordStealingHackingWiFi2023}.

\subsection{The Role of Adversarial Learning}
\label{sec:adaptation_methods}
\begin{figure}[t]
    \centering
    \includegraphics[width=0.6\linewidth]{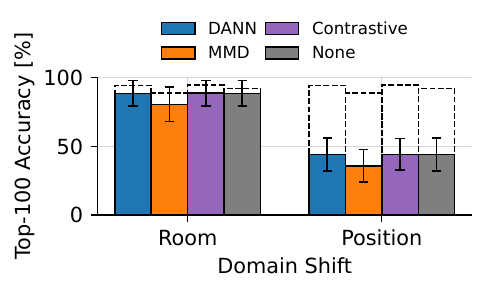}
    \caption{Top-100 accuracy of WiKI-Eve+ under room and position shift when trained with different domain generalization methods (DANN, MMD, contrastive learning) and without an explicit one (None). Error bars show standard deviation across leave-one-out folds and dashed lines show the average in-domain validation accuracy.}
    \label{fig:adaptation_methods}
\end{figure}
Having established WiKI-Eve+ as the strongest realization of the attack, we next assess the mechanism to which WiKI-Eve attributes its cross-domain performance: adversarial learning for domain generalization. To this end, we train WiKI-Eve+ with three different domain objectives -- DANN (default), MMD, and contrastive learning -- each operating on the explicit physical domain labels introduced above, and compare them against an otherwise identical model trained without any domain objective. As shown in~\autoref{fig:adaptation_methods}, MMD yields the lowest performance, while DANN and contrastive learning perform on par with the model trained without an explicit domain objective. Across all objectives the overall trend is the same: strong performance under room shift and a pronounced drop under position shift. These results call into question the central design claim of WiKI-Eve, which presents adversarial learning  as the key enabler of cross-domain generalization. Explicit domain objectives provide no benefit over training without them, suggesting that the domain diversity inherent in the training data already induces a degree of implicit generalization, consistent with prior findings in domain generalization~\cite{gulrajani2020searchlostdomaingeneralization}.

\subsection{Benchmarking WiKI-Eve+ Against the State of the Art}

Having established WiKI-Eve+ as the strongest realization of the WiKI-Eve
approach we could construct, we now compare it against additional
state-of-the-art attacks, WINK~\cite{yangWINKWirelessInference2022} and
WindTalker~\cite{windtalker}. \schemename subjects every method to identical,
controlled domain shifts.

\paragraph{WINK} This attack~\cite{yangWINKWirelessInference2022} is designed to
operate without domain-specific training data. It rests on two assumptions: that
individual keystrokes produce distinctive physical-layer signatures, and that
inter-keystroke timing is a reliable side channel. From the resulting digit
uniqueness and relative distances, the digits typed on a standard PIN pad can be
inferred. We reproduce WINK under idealized conditions to isolate its core
mechanism, assuming perfect keystroke detection and constant typing speed, and
thus exact timing. Where the original work uses dynamic time warping to assess
keystroke similarity, we use the Euclidean distance between \ac{BFI} time
samples because our sequences are already time-aligned.

\paragraph{WindTalker} This baseline~\cite{windtalker} assumes access to a labeled
training set with representative samples per digit and, at inference time,
compares unseen samples against it using a similarity metric, predicting the
class with the lowest average distance to its samples. We instead take the
minimum rather than the average distance per class, which performed best in our
setting. As a template-matching approach, WindTalker is not
designed for domain generalization, and the original work acknowledges sensitivity to environmental conditions.

\begin{figure}[t]
    \centering
    \includegraphics[width=0.65\linewidth]{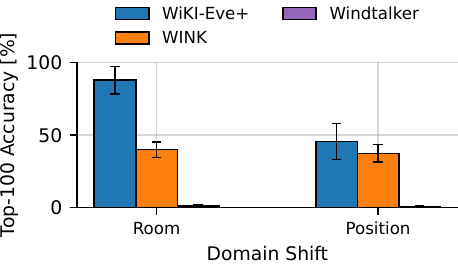}
    \caption{Top-100 accuracy of WiKI-Eve+, WINK, and WindTalker under room and position shift. WiKI-Eve+ outperforms both baselines under identical controlled domain shifts. Error bars show standard deviation across leave-one-out folds.}
    \label{fig:model_comparison}
\end{figure}

\autoref{fig:model_comparison} reports Top-100 accuracy under room and
position shift. WiKI-Eve+ achieves the highest accuracy under both. Under
room shift, WiKI-Eve+ and WINK roughly match the numbers reported in their
original papers: WINK reaches around \SI{40}{\percent}, close to the
\SI{50}{\percent} reported originally~\cite{yangWINKWirelessInference2022}, and
WiKI-Eve+ reaches \SI{88}{\percent} and thus even exceeds the \SI{75}{\percent} reported for WiKI-Eve~\cite{huPasswordStealingHackingWiFi2023}. This changes under
position shift, where WiKI-Eve+ drops to around \SI{46}{\percent}, only
marginally ahead of WINK. WindTalker fails almost entirely under both shifts,
below \SI{1}{\percent}, consistent with a template-matching approach that cannot
bridge a domain gap.\footnote{This is in line with
\autoref{fig:windtalker_performance} in the Appendix, showing that added training
rooms leave out-of-domain accuracy largely flat.} WiKI-Eve+ is thus the
strongest of the three attacks and a clear advance under background-type room shift. 

\section{Evaluation of Domain Shifts}
\label{sec:domain_eval}

We now characterize when \mbox{Wi-Fi}-based PIN inference with the WiKI-Eve+ attack generalizes under \schemename
domain shift and when it fails. We first compare background- and encoding-type
shifts, then dissect the variation within each shift type down to the digit
level, and finally isolate the data properties that enable generalization at
all. Our findings delineate a narrow attacker regime: generalization holds when
only the background changes, but collapses once the encoding of the typing event
is altered. %

\begin{figure}[t]
    \centering
    \includegraphics[width=0.65\linewidth]{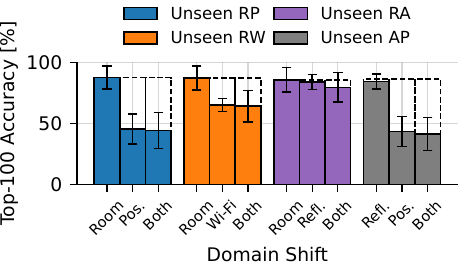}
    \caption{Performance of our WiKI-Eve+ digit prediction model. Per-split
    Top-100 accuracy across all second-order leave-out configurations
    (\cf~\autoref{tab:second-order-leave-out}).}
    \label{fig:combined_transfers}
\end{figure}

\subsection{Comparing Background and Encoding Shifts}

\autoref{fig:combined_transfers} reports both
first-order and second-order test results for each leave-out configuration.
Background-type shifts outperform encoding-type shifts across all settings,
including both single-factor and combined background shifts. This aligns with our
earlier results, which held independent of model architecture and feature
pipeline. The similar results for room and reflector shift suggest that both
types of background shift pose comparable difficulty. For an attacker, this means
relocating to a new room or tolerating rearranged surroundings costs little, and
the background need not be known in advance.

For shifts involving encoding factors (RP, RW, and AP), the dual-factor results
are largely driven by the encoding component. As our theoretical model predicted,
this confirms experimentally that encoding-type variation always entails
concurrent background variation. Interestingly, Wi-Fi channel shift consistently
outperforms position transfer, which we attribute to channel reconfiguration not
involving changes in physical geometry. On closer inspection, channel pairs with
smaller frequency gaps tend to generalize better. While the limited
number of Wi-Fi channels can be enumerated and learned, positions are continuous
and cannot be fully covered during training. A lack of generalization across
position shift thus limits the real-world threat of PIN inference to fixed
attacker-victim geometries.
Notably, although the leave-out tests vary in dataset size by design, size alone
shows no clear effect on performance: the Wi-Fi encoding-shift models were
trained on less data than the position encoding-shift models, yet still
outperform them, ruling out dataset size as the source of this gap.

\begin{figure}[t]
    \centering
    \begin{subfigure}[t]{0.45\linewidth}
        \centering
        \includegraphics[width=\linewidth]{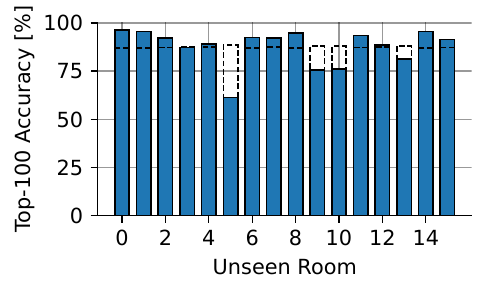}
        \caption{}
        \label{fig:subfig_a}
    \end{subfigure}
    \hfill
    \begin{subfigure}[t]{0.45\linewidth}
        \centering
        \includegraphics[width=\linewidth]{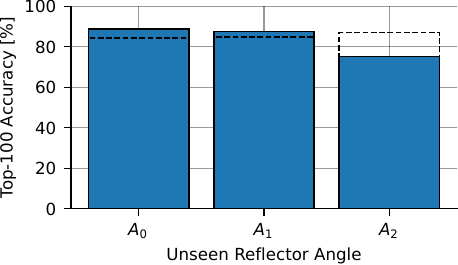}
        \caption{}
        \label{fig:subfig_d}
    \end{subfigure}
    \begin{subfigure}[t]{0.45\linewidth}
        \centering
        \includegraphics[width=\linewidth]{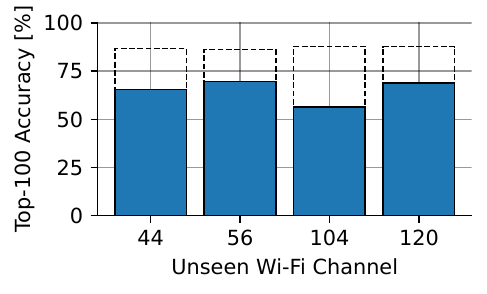}
        \caption{}
        \label{fig:subfig_b}
    \end{subfigure}
    \hfill
    \begin{subfigure}[t]{0.45\linewidth}
        \centering
        \includegraphics[width=\linewidth]{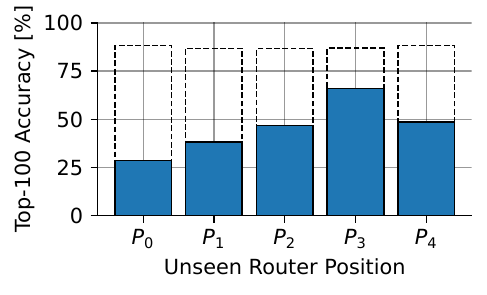}
        \caption{}
        \label{fig:subfig_c}
    \end{subfigure}
    \hfill

    \caption{Generalization performance across individual held-out domain
    factors: (a) rooms, (b) reflector angles, (c) Wi-Fi channels, and (d) router
    positions. Dashed lines show the in-domain validation accuracy for
    reference.}
    \label{fig:per_transfer_factors}
\end{figure}

\subsection{Within-Shift Variation}
\label{sec:deepdive}
We further analyze the variation of individual instances across the leave-out
combinations, over which the standard deviation in
\autoref{fig:combined_transfers} is computed. \autoref{fig:per_transfer_factors}
reports accuracy broken down by individual instances of each domain factor.
Across rooms, the model consistently achieves above \SI{75}{\percent} accuracy,
except for room~5, where performance drops to around \SI{60}{\percent}, which may
reflect its being the smallest environment in the dataset. In contrast, neither
Wi-Fi channel nor reflector angle exhibits consistent instance-specific trends.
For router positions (\autoref{fig:subfig_c}), the highest accuracy is reached at $P_3$ where the smartphone and router are aligned along a straight line, suggesting
that certain geometric configurations may favor generalization, though the
underlying cause requires further investigation.

\begin{figure}[t]
    \centering
    \includegraphics[width=0.75\linewidth]{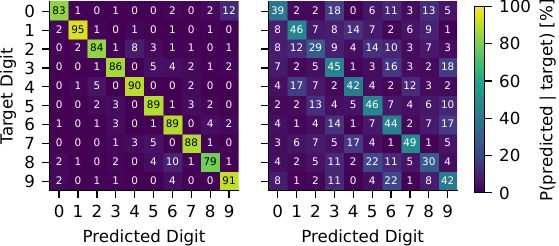}
    \caption{Per-digit confusion matrices on an unseen room (left) and an unseen router position (right).}
    \label{fig:confusion_matrices}
\end{figure}

Finally, we examine two single-domain transfers at the digit level.
\autoref{fig:confusion_matrices} shows per-digit confusion matrices for Room~7
and position~$P_0$, both held out during training. The room transfer achieves an
average digit accuracy of \SI{87}{\percent}, giving a strong diagonal. Its most
prominent off-diagonal error is between 0 and 9, with \SI{12}{\percent} of zeros
misclassified as 9. This confusion is not consistent with spatial proximity on
the keypad, suggesting that errors in background-type transfer are not driven by
geometric digit layout alone. The position transfer achieves substantially lower
accuracy (\SI{41}{\percent}) and a weaker diagonal, with confusions occurring
predominantly between vertically adjacent digits. As the phone is rotated
\SI{90}{\degree} relative to the router, this direction aligns with the axis
along which the router moves during a position transfer. Such structure suggests
that encoding shifts affect \ac{BFI} data in a more organized way than simple
noise.

\subsection{Factors Driving Domain Generalization}

The previous section shows that our deep-learning model achieves a degree of
domain generalization, yet the underlying causes remain unclear. To identify
which factors contribute, we perform a series of ablations that isolate the
impact of specific data properties.

\begin{figure}[t]
    \centering
    \begin{subfigure}[t]{0.45\linewidth}
        \centering
        \includegraphics[width=\linewidth]{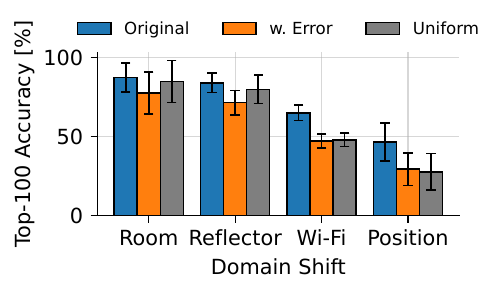}
        \caption{Keystroke Timing}
        \label{fig:timing}
    \end{subfigure}
    \hfill
    \begin{subfigure}[t]{0.45\linewidth}
        \centering
        \includegraphics[width=\linewidth]{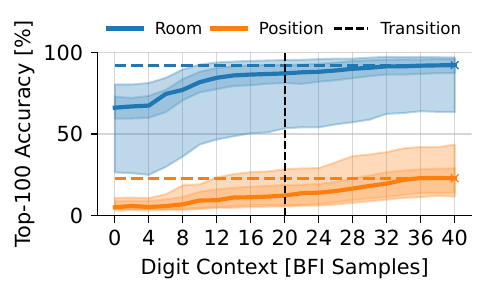}
        \caption{Digit Context}
        \label{fig:context_size}
    \end{subfigure}
    \caption{Effect of data properties on PIN inference.}
    \label{fig:timing_context}
\end{figure}

Key features in \ac{BFI} sequences are closely tied to the typing process,
particularly inter-keystroke timing (exploited by
WINK~\cite{yangWINKWirelessInference2022}) and inter-keystroke transients
(leveraged by WiKI-Eve~\cite{huPasswordStealingHackingWiFi2023}). To assess
whether our models rely on these signals, we perturb timing and remove
neighboring-digit context.

\autoref{fig:timing} shows the effect of timing across domain shifts. The
baseline (from \autoref{fig:combined_transfers}) assumes constant typing speed.
Adding random timing noise (Gaussian, $\sigma=3$ samples) reduces accuracy across
all domains, indicating partial reliance on timing. Under enforced uniform
timing, which removes the timing side channel, background transfers (room,
reflector) largely recover while encoding transfers (channel, position) remain
degraded. Timing is thus not essential for generalization across background
transfers, but can improve performance under encoding variation.

We next analyze the role of neighboring-digit context under the uniform-timing
setting, varying the context window from 0 to 40 samples, i.e., from a single
sample per keystroke to segments extending to the adjacent keystrokes
(\autoref{fig:context_size}). Even without context, a single sample per
keystroke, background transfer reaches around \SI{66}{\percent} accuracy whereas
position transfer largely fails. Larger context improves both, indicating that
short temporal windows already capture informative motion cues. As the window
begins to include transients toward neighboring keystrokes, accuracy approaches
the full-context model, with diminishing returns beyond that point.

\section{Limit Exploration}

The previous sections show that domain generalization works well for background-only transfers but degrades once encoding variation is introduced, an important factor for attacks under device mobility. We now test the limits of this generalization beyond the physical domains supported by our typing station. Unless stated otherwise, we use a model trained on the full \schemename dataset and evaluate it on additional transfer experiments. We find that generalization performance holds in a neighborhood of the trained configuration but collapses once the physical scene changes more substantially, \eg, when device distance, router angle, or smartphone model change. An attacker who can anticipate the target scene may therefore train for it, but generalizing across arbitrary conditions appears difficult even with the rich domain diversity in our training data.

\subsection{External Influences}

\begin{figure}[t]
    \centering
    \begin{subfigure}[t]{0.45\linewidth}
        \centering
        \includegraphics[width=\linewidth]{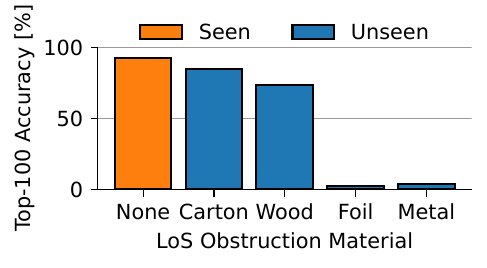}
        \caption{}
        \label{fig:obstruction}
    \end{subfigure}
    \hfill
    \begin{subfigure}[t]{0.45\linewidth}
        \centering
        \includegraphics[width=\linewidth]{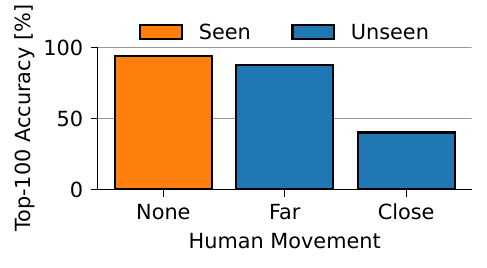}
        \caption{}
        \label{fig:walking}
    \end{subfigure}
    \caption{Effect of external influences during the attack: (a) line-of-sight obstruction, (b) human movement in the victim environment.}
    \label{fig:external_factors}
\end{figure}

\paragraph{Line-of-sight obstruction}
Obstructing the line of sight between phone
and router induces an encoding-type shift by altering the propagation paths from
router to hand. \autoref{fig:obstruction} reports Top-100 accuracy with no
obstruction and with cardboard, wood, aluminum foil, and a metal plate.
Low-attenuation materials cause only minor degradation, whereas highly
reflective or blocking materials, such as aluminum foil and metal, reduce accuracy.

\paragraph{Background movement} 
Where all preceding experiments kept the room
static apart from the typing, we now have a person walk continuously either
behind the router (\emph{far}) or beside the phone (\emph{close}).
\autoref{fig:walking} shows that distant movement has little effect, whereas
movement near the phone substantially degrades accuracy. Such nearby motion
induces strong, dynamic channel variation, a dynamic background-type shift the
model, trained only on static shifts, does not handle.

\subsection{Router Translations}

\begin{figure}[t]
    \centering
    \begin{subfigure}[t]{0.45\linewidth}
        \centering
        \includegraphics[width=\linewidth]{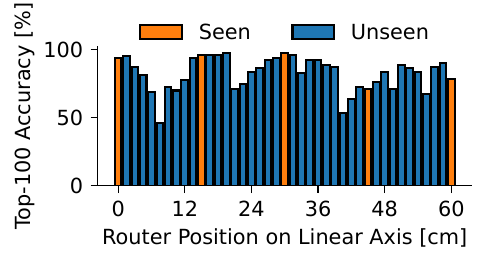}
        \caption{}
        \label{fig:router_axis_position}
    \end{subfigure}
    \hfill
    \begin{subfigure}[t]{0.45\linewidth}
        \centering
        \includegraphics[width=\linewidth]{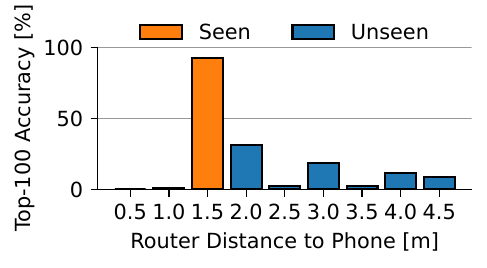}
        \caption{}
        \label{fig:router_distance}
    \end{subfigure}
    \begin{subfigure}[t]{0.45\linewidth}
        \centering
        \includegraphics[width=\linewidth]{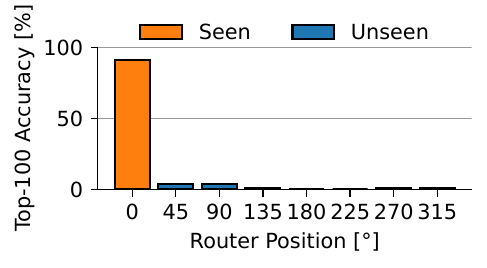}
        \caption{}
        \label{fig:router_position}
    \end{subfigure}
    \hfill
    \begin{subfigure}[t]{0.45\linewidth}
        \centering
        \includegraphics[width=\linewidth]{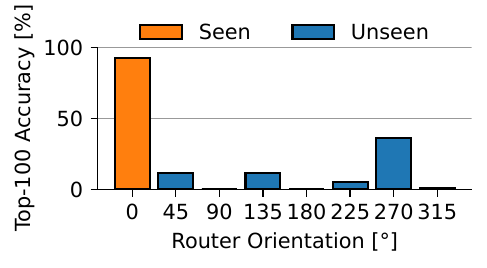}
        \caption{}
        \label{fig:router_rotation}
    \end{subfigure}
    \caption{Generalization across additional router deployment conditions. (a) Intermediate positions along the linear axis. (b) Variation of router–phone distance. (c) Angular router positions around the phone at fixed distance and orientation. (d) In-place router rotation at a fixed location.}
    \label{fig:router_factors}
\end{figure}

\paragraph{Intermediate axis positions}
Our earlier results show limited generalization across router positions spaced \SI{15}{\cm} apart, with no leave-out setting exceeding \SI{60}{\percent} Top-100 accuracy. To test whether this limitation persists at finer granularity, we evaluate the full-dataset model at intermediate positions spaced \SI{1.5}{\cm} apart. As shown in \autoref{fig:router_axis_position}, performance stays above \SI{50}{\percent} except at one position, though positions midway between two
training points show slight degradation. Generalization thus holds along the observed linear axis, but, as we show next, does not extend to more general changes in router placement.

\paragraph{Distance} 
We next vary the distance between router and phone. As shown in
\autoref{fig:router_distance}, accuracy drops markedly as the distance deviates from the training setup, both when the router moves closer and farther away, yet at \SI{2}{\m} and \SI{3}{\m} the model still reaches \SI{30}{\percent} and \SI{18}{\percent} Top-100 accuracy. This exposes a key limitation: generalization does not transfer uniformly across related physical dimensions. Although position
and distance are closely related, robustness along one axis does not imply
robustness along the other. This contrasts with the results reported for
WiKI-Eve~\cite{huPasswordStealingHackingWiFi2023}, where accuracy degrades only gradually with distance.

\paragraph{Orientation and relative positioning} 
We further examine generalization under changes in router placement and
orientation. In the first experiment, the router moves around the phone at
varying angular positions while distance and orientation are held constant. In the second, the router stays in place but rotates about its own axis. In both, the router-phone distance is fixed at \SI{1.5}{\m}. As shown in
\autoref{fig:router_factors}, the model fails to generalize under either
variation. Under rotation (\autoref{fig:router_rotation}), no consistent pattern emerges. Under angular placement (\autoref{fig:router_position}), accuracy improves slightly as the router approaches its training position, indicating that the learned representation is local.

\subsection{Phone-Related Shift}

As the final class of experiments, we consider changes to the typing and the smartphone. %

\begin{figure}[t]
    \centering

    \begin{subfigure}[t]{0.45\linewidth}
        \centering
        \includegraphics[width=\linewidth]{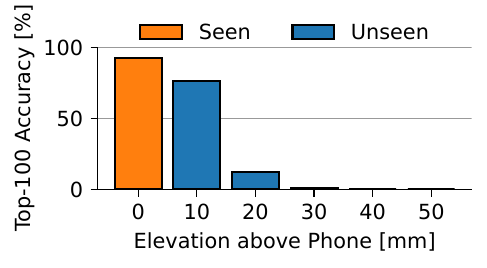}
        \caption{}
        \label{fig:elevation}
    \end{subfigure}
    \hfill
    \begin{subfigure}[t]{0.45\linewidth}
        \centering
        \includegraphics[width=\linewidth]{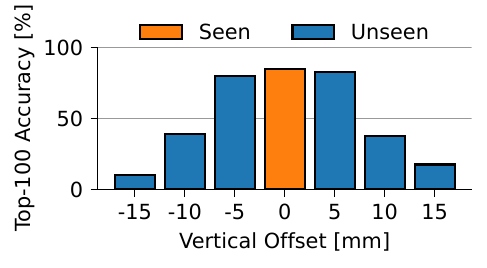}
        \caption{}
        \label{fig:yoffset}
    \end{subfigure}
    \begin{subfigure}[t]{0.45\linewidth}
        \centering
        \includegraphics[width=\linewidth]{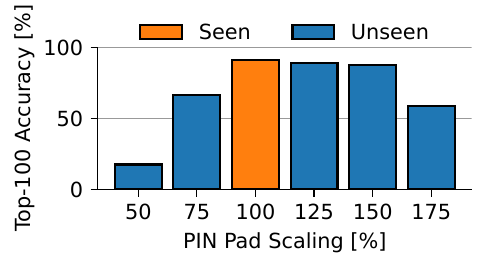}
        \caption{}
        \label{fig:scaling}
    \end{subfigure}
    \hfill
    \begin{subfigure}[t]{0.45\linewidth}
        \centering
        \includegraphics[width=\linewidth]{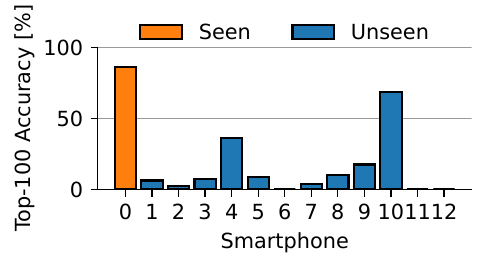}
        \caption{}
        \label{fig:phones}
    \end{subfigure}
    \caption{Attack generalization under phone-related variations: (a) elevated typing, (b) vertical offset, (c) PIN pad scaling, (d) smartphone models.}
    \label{fig:scaling_and_phone}
\end{figure}

\paragraph{PIN pad position and scaling}
We evaluate robustness to shifts in the PIN pad location by typing with spatial
offsets via the robotic arm. In the first experiment, the typing plane is raised above the phone. In the second, the PIN pad is shifted vertically on the display, corresponding to a countermeasure that relocates the keypad. Both induce an encoding-only shift, affecting $H_\mathrm{h}(t)$ while leaving $H_\mathrm{d}$ and $H_\mathrm{e}$ unchanged. \autoref{fig:scaling_and_phone}a and~b show
that the model generalizes to offsets up to \SI{10}{\mm} but degrades beyond. Given that a small keypad displacement substantially disrupts the attack, randomizing the keypad position could serve as a countermeasure. We then vary the scale of the PIN pad,
simulating a larger or smaller pad while keeping antenna placement fixed.
\autoref{fig:scaling} shows that Top-100 accuracy stays above \SI{50}{\percent} for scaling factors between \SI{75}{\percent} and \SI{175}{\percent}, suggesting that the model relies on relative features, such as the direction of finger movement, rather than an absolute mapping of channel responses to digit positions.

\paragraph{Phones}
Finally, we evaluate generalization across 12 additional phones, listed
in~\autoref{tab:phones} in the Appendix. Besides a second iPhone~13, this set spans a range of commodity smartphones and includes all but one of the devices reported in WiKI-Eve~\cite{huPasswordStealingHackingWiFi2023}.
\autoref{fig:phones} shows the results. Top-100 accuracy stays below
\SI{10}{\percent} for nine devices, reaching around \SI{20}{\percent} for phone~9
(a Samsung Galaxy~S23) and roughly \SI{70}{\percent} for phone~10 (a Samsung
Galaxy~S25 Ultra). Generalization to unseen phones is thus possible but
inconsistent and device-dependent: even a second iPhone~13, identical in model
to the training device, reaches only around \SI{40}{\percent}.

\section{Discussion}
\label{sec:discussion}

In this section, we discuss the experimental setup, limitations, our results and reflect on threat potential.
Finally, we reason about countermeasures and provide directions for future research.

\paragraph{Experimental setup}
Our experiments assume that the router periodically requests \ac{BFI} from the victim smartphone. While some other \mbox{Wi-Fi} routers did not issue such requests despite transmit beamforming being enabled, the router used in our experiments does so under sufficient traffic, which we induce via ICMP ping packets. Even when \ac{BFI} is not natively available, an attacker can obtain channel estimates through alternative means such as passively estimating \ac{CSI} from observed \mbox{Wi-Fi} frames. The latter can succeed even when the victim is unassociated, by injecting frames that trigger NIC acknowledgments~\cite{abediWiFiPhysicalLayer2024}. Data collection requires approximately 24~hours per room and took place in a typical office building. During this time, incidental human motion in adjacent rooms or corridors and slow environmental drift (\eg, temperature or humidity changes) may introduce minor variation. While such effects would fall within the scope of our threat model, we did not observe any such impact on the recorded data.

\paragraph{Limitations}
While our setup separates encoding from background shifts, it can only produce
the physical domain shifts the typing station supports. The \ac{EM} phantom hand
approximates but does not fully replicate a human hand~\cite{hand_phantom_effects_2009}: it remains static during typing and reproduces neither fine-grained muscular motion nor natural variation
in hand-to-device positioning. The robotic arm likewise enforces a single fixed
typing style, whereas human typing exhibits inherent variation such as one- versus two-handed typing. These simplifications mean our absolute accuracies describe an idealized victim, not a human one, and the model is not meant to generalize to human typing. In the context of our domain-shift
analysis, however, they act in the attacker's favor, removing behavioral noise that would otherwise hinder inference. Our results thus describe a best case for
the attacker: a domain shift that already breaks the attack under these idealized conditions will break it at least as much against a human, whose typing adds
further variation.

\paragraph{Results and takeaways}

Our evaluation yields three takeaways.

First, domain shifts fall into distinct classes. Background
changes, such as a new room or rearranged surroundings, leave the typing signal
intact and are readily overcome, with Top-100 accuracies above \SI{80}{\percent}.
Encoding changes, such as a new device position or a different Wi-Fi channel,
reshape the typing signal itself and cause substantial degradation, with Top-100 accuracies below \SI{50}{\percent}. This split, predicted by our channel model, held for every architecture, feature pipeline, and domain generalization objective tested.

Second, this limitation is not an artifact of a weak attacker. WiKI-Eve+
outperforms every reproduction and baseline we constructed, and we grant it
near-ideal conditions: perfectly repeatable typing, perfect segmentation, and
exact keystroke timing. Even so, no configuration generalizes reliably across
encoding shifts. A real attacker, facing human typing variability and imperfect
segmentation, can only do worse.

Third, the difficulty of a domain shift is not apparent from its description. Two
rooms can differ visibly yet yield similar keystroke signals, while a few
centimeters of router displacement can render them unrelated. A single successful
transfer therefore says little: assessing generalization requires averaging over
many instances of a shift, and generalization along one axis does not imply it
along another. Accuracies reported on a few hand-picked conditions cannot
substantiate a general threat.

Together, these findings show that cross-domain generalization in Wi-Fi sensing
remains a genuinely difficult and unsolved problem. On our controlled benchmark,
current attacks pose a threat only where the attacker-victim geometry is
fixed, \eg, at a \ac{POS} terminal. In the scenario that would make the attack a
broad threat, a mobile target in an unfamiliar location, we did not observe
generalization. We do not adjudicate the absolute accuracies reported in prior
work. Instead, we release \schemename as a public, controlled benchmark on which
such claims can be examined under identical conditions, and we invite the
community to test whether stronger generalization is achievable where our attacks
fall short.

\paragraph{Countermeasures}
Wi-Fi-based PIN inference can be counteracted at the application,
protocol, or physical layer. At the application level, PIN-based authentication
can be replaced with biometric methods, or the on-screen keypad layout
randomized to disrupt the spatial correlations the attack exploits, though such measures may cost usability
or be unavailable on devices without biometric support. At the protocol level,
\ac{BFI} transmission could be disabled or encrypted, but this addresses only one
symptom: the threat arises from the physical observability of \mbox{Wi-Fi}
signals, so an attacker can fall back on independently estimating \ac{CSI} from
ambient traffic. Protocol-level defenses are therefore insufficient on their own.
The most robust mitigations act at the physical layer, altering or obfuscating
the channel observations themselves, for instance through reconfigurable surfaces
that distort propagation paths~\cite{irshield} or transmission schemes that
prevent unauthorized decoding~\cite{luoMIMOCryptMultiUserPrivacyPreserving2024}.

\paragraph{Future work}
\schemename enables several directions for future work. The robotic platform can
be extended to a broader range of typing styles and physical domains, and paired
robot-human studies could quantify residual physiological variability and assess
transfer to human typing. A further direction is the adversarial replication of a
victim's environment to collect targeted training data, which may enable more
reliable encoding transfer. Beyond \mbox{Wi-Fi}, PIN inference should also be examined for other wireless standards such as 5G and Bluetooth. More fundamentally, the controlled-actuation principle carries over to other wireless sensing tasks, such
as gestures, gait, and facial expressions, providing a basis for systematic
threat assessment of RF-based inference attacks. Finally, our dataset may support
generative models that synthesize multi-domain data, easing the collection burden
for future studies.

\section{Related Work}

Early work on Wi-Fi-based keystroke inference operates under white-box assumptions. \mbox{WiKey}~\cite{aliKeystrokeRecognitionUsing2015, aliRecognizingKeystrokesUsing2017} demonstrated keystroke recognition from \ac{CSI} under user-specific training and controlled conditions. Subsequent work extended this to smartphones and touch input~\cite{zhangPrivacyLeakageMobile2016, shenWiPass1DCNNbasedSmartphone2021}, including \mbox{WindTalker}~\cite{liWhenCSIMeets2016, windtalker}, which likewise pursues a template-matching approach. More recent approaches claim some degree of domain generalization, an open challenge in the general wireless sensing context~\cite{chenCrossDomainWiFiSensing2023}. WINK~\cite{yangWINKWirelessInference2022} eliminates explicit training by matching recurring \ac{CSI} patterns and inter-keystroke timing against a known keypad layout, conceptually similar to linguistic-structure-based approaches for keyboards~\cite{fangNoTrainingHurdles2018, yangWirelessTrainingFreeKeystroke2022}. WiKI-Eve~\cite{huPasswordStealingHackingWiFi2023} reports high PIN inference accuracy from \ac{BFI} and attributes cross-domain performance to adversarial learning, with subsequent extensions to more complex scenarios~\cite{wangMuKIFiMultiPersonKeystroke2024, chenSilentThiefPassword2024, chenEchoesFingertipUnveiling2024}. More broadly, adversarial wireless sensing, \eg, using mmWave radar, has been shown to expose sensitive information beyond keystrokes, including motion~\cite{zhuTuAlexaWhen2020, xiaoLendMeYour2025}, voice~\cite{basakMmSpySpyingPhone2022}, and handwriting~\cite{shichenRadSeeSeeYour2025}. At the same time, wireless sensing systems can be attacked through spoofing~\cite{jiangRISirenWirelessSensing2024, yizhuTileMaskPassiveReflectionbasedAttack2023, reddyvennamMmSpoofResilientSpoofing2023}, while defenses explore obfuscation~\cite{irshield} and physical-layer encryption~\cite{luoMIMOCryptMultiUserPrivacyPreserving2024}.

\section{Conclusion}

We introduced \schemename, a methodology for the systematic threat assessment of
Wi-Fi-based PIN inference. By automating PIN entry with a robotic arm, it makes
the sensing event reproducible and, for the first time, attributes changes in
attack performance to individual physical conditions, exposing the fundamental
challenges behind cross-domain generalization. Guided by a channel model that
separates domain shifts into background and encoding changes, we collect a
large-scale dataset of over one million typed digits across more than one
thousand controlled conditions. Our evaluation shows that current attacks
generalize across background changes but degrade substantially once the encoding
of the typing itself is altered, precisely the condition a real attacker cannot
control. This holds even for a deliberately strengthened attacker under idealized
conditions, and the adversarial learning credited for prior cross-domain results
yields no consistent benefit. The practical threat is therefore far narrower than
single-axis evaluations suggest, confined to settings with fixed
attacker-victim geometry such as a \ac{POS} terminal. Cross-domain
generalization in Wi-Fi sensing thus remains an open problem. We release
\schemename as the controlled basis on which such claims can be examined, and
invite the community to test whether stronger generalization is achievable where
current attacks fall short.

\section*{Acknowledgment}
We thank Maximilian Andersen, Matej Grado\v{s} and \mbox{Sebastian} Caroli for their support during initial attack exploration. This work was in part funded by the Deutsche Forschungsgemeinschaft (DFG, German Research Foundation) under Germany's Excellence Strategy - EXC 2092 CASA - 390781972..

\bibliographystyle{IEEEtran}
\bibliography{wspinbib}

\appendix

\subsection{Features}

\begin{table}[t]
\centering
\footnotesize
\renewcommand{\arraystretch}{1.2}
\setlength{\tabcolsep}{4pt}
\begin{tabular}{@{}r r l r c p{4.0cm}@{}}
\toprule
\textbf{\#ID} & \textbf{Index} & \textbf{Name} & \textbf{\#} & \textbf{Data} &\textbf{Description} \\
\midrule
1  & 0--7     & \texttt{nAmp}  & 8  & $\tilde{V}$ & Amplitude, averaged over subcarriers per channel. \\
2  & 8--13    & \texttt{nPhs}  & 6  & $\tilde{V}$ & Unwrapped phase, averaged over subcarriers per channel. \\
3  & 14--23   & \texttt{nAng}  & 10 & $\hat{V}$ & Raw compressed angles, DC-normalized, averaged over subcarriers, per angle.\\
4  & 24--31   & \texttt{ed0}    & 8  & $V$ &  Euclidean distance to the first sample, per channel. \\
5  & 32--39   & \texttt{edR}    & 8  & $V$ &  Euclidean distance to $\tilde{V}_{\text{Ref}}$, per channel. \\
6  & 40--42   & \texttt{gR}    & 3  & $V$ &  Grassmann geodesic distance between consecutive samples, to the first sample and to $\tilde{V}_{\text{Ref}}$. \\
7  & 43--45   & \texttt{g}   & 3  & $\tilde{V}$ &  Same three Grassmann distances, computed on $V$. \\
8  & 46--61   & \texttt{dfs}   & 16 & $V$ &  Spectrogram magnitude in the lower and upper band, per channel. \\
9  & 62--69   & \texttt{mrc}   & 8  & $V$ &  Maximum-ratio combined amplitude, DC-normalized, per channel. \\
10 & 70--77   & \texttt{hAmp}    & 8  & $V$ &  Mean of the 5 highest-variance amplitude subcarriers, per channel. \\
11 & 78--83   & \texttt{hPhs}    & 6  & $V$ &  Mean of the 5 highest-variance unwrapped phase subcarriers, per channel. \\
12 & 84--91   & \texttt{lAmp}    & 8  & $V$ &  Mean of the 5 lowest-variance amplitude subcarriers, per channel. \\
13 & 92--97   & \texttt{lPhs}    & 6  & $V$ &  Mean of the 5 lowest-variance unwrapped phase subcarriers, per channel. \\
14 & 98--107  & \texttt{pcaAng}    & 10 & $\hat{V}$ &  First principal component of each compressed angle. \\
15 & 108--115 & \texttt{pcaAmp}  & 8  & $\tilde{V}$ &  First principal component of amplitude, per channel. \\
16 & 116--121 & \texttt{pcaPhs}  & 6  & $\tilde{V}$ &  First principal component of unwrapped phase, per channel. \\
17 & 122--133 & \texttt{steer}    & 12  & $\tilde{V}$ &  Inter-TX-element phase differences for stream 0 and 1 (6 TX antenna pairs each). \\
\bottomrule
\end{tabular}
\caption{Our proposed feature catalogue containing 134 features, extracted from the \ac{BFI} timeseries grouped into 17 feature-classes. Features using the phase have only 6 channels, because the phase of the 7th and 8th channel of \ac{BFI} is zero by design.}
\label{tab:features}
\end{table}

\ac{BFI} timeseries are inherently high dimensional as channel information is estimated across multiple subcarrier frequencies and potentially also multiple transmitter-receiver antenna pairs. 
As outlined before, our setup utilizes \SI{80}{MHz} of bandwidth and therefore $N_S = 234$ subcarriers as well as $N_{T} = 4$ transmit and $N_{R} = 2$ receive antennas resulting in $N_A=10$ beamforming angles. This matrix $\hat{V}^{N_S \times N_A}$, \ie, $2340$ integer-valued angles are transmitted in the compressed beamforming report of each Action-No-Ack \mbox{Wi-Fi} management frame. Our pipeline gathers those matrices across~$T$ time samples and decompresses them into the complex-valued \ac{BFI} channel matrix $\tilde{V}^{T \times N_S \times N_{T} \times N_{R}}$.

From $\tilde{V}$, we compute a normalized \ac{BFI} timeseries $V^{T \times N_S \times N_{T} \times N_{R}}$ that relies on a single reference \ac{BFI} sample $\tilde{V}_{\text{Ref}}$. The features are then derived from the raw beamforming angle timeseries $\hat{V}$, the complex-valued timeseries $\tilde{V}$ and the complex-valued reference-normalized timeseries $V$. Based on the 17~feature classes listed in~\autoref{tab:features}, we obtain a feature vector $F^{T\times134}$ that contains $134$ features per time instance.

\subsection{Model Architecture}
\begin{figure}[t]
    \centering
    \includegraphics[width=\linewidth]{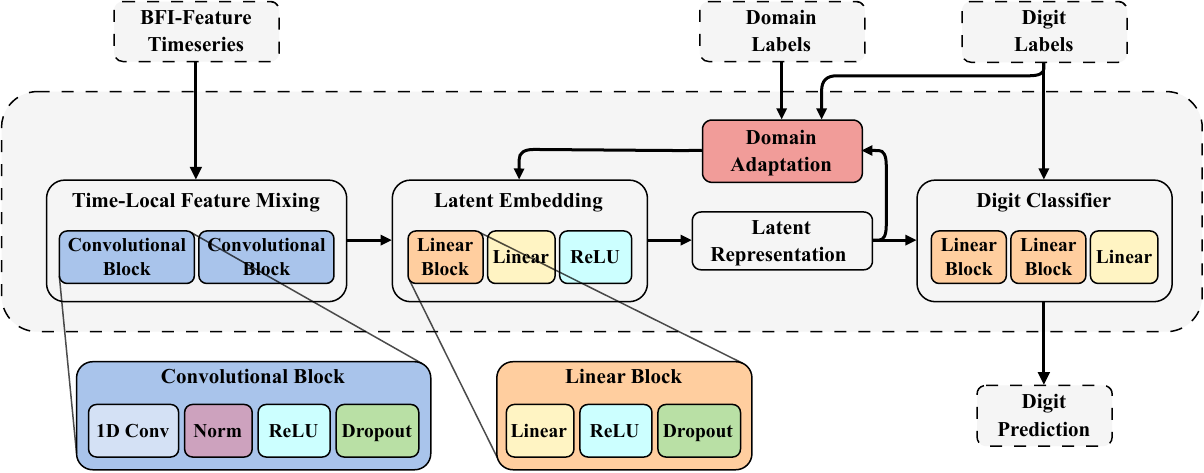}
    \caption{Architecture of our WiKI-Eve+ digit prediction model.}
    \label{fig:model_arch}
\end{figure}

In \autoref{fig:model_arch} the architecture of our WiKI-Eve+ digit prediction model can be seen. The main modifications compared to the original variant~\cite{huPasswordStealingHackingWiFi2023} are the Time-Local Feature Mixing and the Latent Embedding, which give the model more capacity to extract domain invariant features. Our framework further supports multiple domain generalization objectives besides DANN, including MMD~\cite{long15} and contrastive alignment~\cite{Khosla_NEURIPS2020}. The Python implementation can be found in \cite{pinsight_repo}.

\subsection{Model Selection}
\begin{figure}
    \centering
    \includegraphics[width=0.75\linewidth]{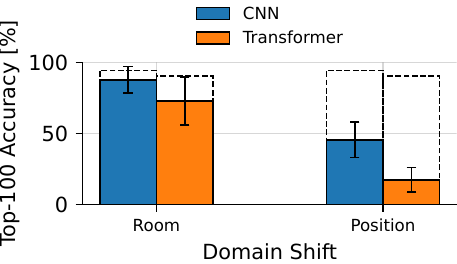}
    \caption{WiKI-Eve+ with a convolutional versus a Transformer feature extractor, under first-order room and position leave-out. Error bars show standard deviation across leave-one-out folds; dashed lines the average in-domain validation accuracy.}
    \label{fig:trafo}
\end{figure}

To guard against an evaluation bias from a fixed model architecture, we implemented several variants of the feature-extraction module within WiKI-Eve+. \autoref{fig:trafo} compares the final convolutional variant against one in which a Transformer encoder operates directly on the input feature timeseries, under a first-order room and position leave-out. Although both reach comparable in-domain accuracy, the convolutional variant generalizes markedly better, reaffirming it as our strongest realization of the attack. Moreover, the general trend of background transfers outperforming encoding transfers persists here, indicating that the observed cross-domain behavior stems from the signal itself rather than from the specific choice of feature extractor.

\subsection{Rooms}
\autoref{tab:rooms} lists descriptions of the rooms used in the experimental study with the indices matching the results shown in \autoref{fig:subfig_a}.

\begin{table}[H]
\centering
\footnotesize
\renewcommand{\arraystretch}{1.2}
\setlength{\tabcolsep}{4pt}
\begin{tabular}{@{}c r p{5.8cm}@{}}
\toprule
\textbf{Index} & \textbf{Size [\SI{}{m\squared}]} & \textbf{Description} \\
\midrule
0, 15 & 35 & Conference room with a single table. \\
1, 8  & 50 & Laboratory environment. \\
2     & 20 & Conference room with one large desk, a stand-up presentation screen and windows on one side. \\
3     & 15 & Office with two desks and windows on one side. \\
4     & 15 & Comparable office room to 3. \\
5     & 3 & Server room. \\
6, 7  & 35 & Conference room with one large desk, a screen and windows on one side. \\
9    & 3 & Small passage between floors with doors on both sides. \\
10    & 25 & Conference room with a large screen besides the setup. \\
11    & 40 & Conference room with glass doors on both sides. \\
12    & 20 & Conference room without windows, in the middle of the building.\\
13    & 15 & Laboratory environment. \\
14    & 100 & Lecture room with multiple rows of tables. \\
\bottomrule
\end{tabular}
\caption{Rooms used in the experiments with approximate sizes and a brief description. Rooms with multiple indices were recorded in separate sessions while the setup was placed in different locations inside the room.}
\label{tab:rooms}
\end{table}

\subsection{Phones}
\autoref{tab:phones} lists the smartphones used in our experimental study. The indices correspond to the performances shown in \autoref{fig:phones}. For the iPhone 13 and the Xiaomi 13T Pro we tested two devices of the same model to assess intra-model device variances.

\begin{table}[H]
\centering
\footnotesize
\renewcommand{\arraystretch}{1.2}
\setlength{\tabcolsep}{4pt}
\begin{tabular}{@{}c p{3.6cm}@{}}
\toprule
\textbf{Index} & \textbf{Description} \\
\midrule
0, 4 & iPhone 13 \\
1 & Google Pixel 6a \\
2     & Huawei P40 Pro \\
3     & iPhone 13 Mini \\
5     & iPhone 15 \\
6  & OnePlus 10T \\
7  & Samsung Galaxy A36 \\
8  & Samsung Galaxy Z Flip6 \\
9    & Samsung Galaxy S23 \\
10    & Samsung Galaxy S25 Ultra \\
11, 12    & Xiaomi 13T Pro \\
\bottomrule
\end{tabular}
\caption{Smartphones used in the experiments. For the phones with two indices separate phones of the same model were used.}
\label{tab:phones}
\end{table}

\subsection{EM Phantom Effect}
\begin{figure}[t]
    \centering
    \includegraphics[width=0.75\linewidth]{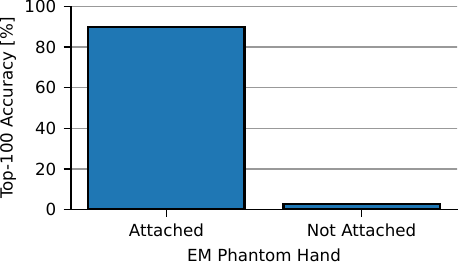}
    \caption{The accuracy of the model drops to \SI{2.5}{\percent} when the EM phantom hand is not attached to the robotic arm.}
    \label{fig:phantom_hand_effect}
\end{figure}

\autoref{fig:phantom_hand_effect} shows the drop in performance when the model is tested without the EM phantom hand being attached to the robotic arm. This shows that typing patterns in $H_h$ originate from the hand rather than from the robotic arm itself.

\subsection{Effect of Domain Diversity on WindTalker}
\begin{figure}[H]
    \centering
    \includegraphics[width=0.75\linewidth]{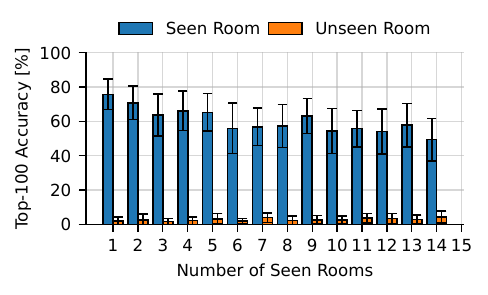}
    \caption{Domain transfer performance of WindTalker. As the number of rooms used for training grows, the in-domain accuracy decreases, while the out-of-domain accuracy stagnates. This shows that just adding more data to the model does not increase domain transfer performance.}
    \label{fig:windtalker_performance}
\end{figure}

To understand the reasons behind pattern matching approaches like WindTalker \cite{windtalker} not working across multiple domains, \autoref{fig:windtalker_performance} shows the accuracy on a seen and on an unseen room as the number of seen rooms during training increases. The performance decreases from above \SI{70}{\percent} to below \SI{50}{\percent} as more rooms are added to the training corpus. At the same time the accuracy on unseen rooms stagnates below \SI{5}{\percent}.

\end{document}